\documentclass[12pt]{article}
\usepackage{graphics,graphicx}
\usepackage{amssymb,epsfig,amsmath,euscript,array}
\usepackage[nosort]{cite}
\usepackage{axodraw4j}
\usepackage{pstricks}
\usepackage{color}

\makeatletter
\@addtoreset{equation}{section}
\makeatother

\newcounter{multieqs}

\newcommand{\be}{\begin{equation}}
\newcommand{\ee}{\end{equation}}

\newcommand{\bm}[1]{\mbox{\boldmath $#1$}}

\newcommand{\kslash}{k \!\!\! / }

\newcommand{\lslash}{l \!\! / }
\newcommand{\Pslash}{P \!\!\!\! / }

\newcommand{\islash}{i \!\!\! / }
\newcommand{\jslash}{j \!\!\! / }
\newcommand{\aslash}{a \!\!\! / }
\newcommand{\bslash}{{b \hspace{-6pt} \slash} }

\newcommand{\onslash}{1 \!\!\! / }
\newcommand{\twslash}{2 \!\!\!/ }
\newcommand{\thslash}{3 \!\!\!/ }
\newcommand{\foslash}{4 \!\!\! / }
\newcommand{\fislash}{5 \!\!\! / }

\newcommand{\mslash}{m \!\!\! / }

\def\bd{\begin{document}}
\def\ed{\end{document}}
\def\nn{\nonumber}
\def\bea{\begin{eqnarray}}
\def\eea{\end{eqnarray}}

\def\ab{(ijab)}
\def\ba{(ijba)}
\def\ijab{{\tr}_{+}(\islash\, \jslash\, \aslash \, \bslash)}
\def\ijba{{\tr}_{+}(\islash\, \jslash\, \bslash \, \aslash)}
\def\ijaP{{\tr}_{+}(\islash\, \jslash\, \aslash \, \Pslash)}
\def\ijPLa{{\tr}_{+}(\islash\, \jslash\, \Pslash_L \, \aslash)}
\def\ijaPL{{\tr}_{+}(\islash\, \jslash\, \aslash \, \Pslash_L)}
\def\ijPLza{{\tr}_{+}(\islash\, \jslash\, \Pslash_{L;z} \, \aslash)}
\def\ijaPLz{{\tr}_{+}(\islash\, \jslash\, \aslash \, \Pslash_{L;z})}
\def\ijPa{{\tr}_{+}(\islash\, \jslash\, \Pslash \, \aslash)}
\def\iaPb{{\tr}_{+}(\islash\, \aslash\, \Pslash \, \bslash)}
\def\ibPa{{\tr}_{+}(\islash\, \bslash\, \Pslash \, \aslash)}
\def\ijPmu{{\tr}_{+}(\islash\, \jslash\, \Pslash \, \mu)}
\def\ibmuP{{\tr}_{+}(\islash\, \bslash\, \mu \, \Pslash)}
\def\ibmua{{\tr}_{+}(\islash\, \bslash\, \mu \, \aslash)}
\def\iamub{{\tr}_{+}(\islash\, \aslash\, \mu \, \bslash)}
\def\jaPb{{\tr}_{+}(\jslash\, \aslash\, \Pslash \, \bslash)}
\def\ijmuP{{\tr}_{+}(\islash\, \jslash\, \mu \, \Pslash)}
\def\ijmum{{\tr}_{+}(\islash\, \jslash\, \mu \, \mslash)}
\def\ijmmu{{\tr}_{+}(\islash\, \jslash\, \mslash \, \mu)}
\def\ijmP{{\tr}_{+}(\islash\, \jslash\, \mslash \, \Pslash)}
\def\iabP{{\tr}_{+}(\islash\, \aslash\, \bslash \, \Pslash)}
\def\ijbP{{\tr}_{+}(\islash\, \jslash\, \bslash \, \Pslash)}
\def\jbPa{{\tr}_{+}(\jslash\, \bslash\, \Pslash \, \aslash)}
\def\ijPb{{\tr}_{+}(\islash\, \jslash\, \Pslash \, \bslash)}
\def\jbmua{{\tr}_{+}(\jslash\, \bslash\, \mu \, \aslash)}

\def\loablt{ {\tr}_{+}(\lslash_1\, \aslash \, \bslash\, \lslash_2)}

\def\ijlolt{{\tr}_{+}(\islash\, \jslash\, \lslash_1 \, \lslash_2)}
\def\ijltlo{{\tr}_{+}(\islash\, \jslash\, \lslash_2 \, \lslash_1)}
\def\ibloa{{\tr}_{+}(\islash\, \bslash\, \lslash_1 \, \aslash)}
\def\jaltb{{\tr}_{+}(\jslash\, \aslash\, \lslash_2 \, \bslash)}
\def\ialtb{{\tr}_{+}(\islash\, \aslash\, \lslash_2 \, \bslash)}
\def\bltloa{{\tr}_{+}(\bslash\, \lslash_2\, \lslash_1 \, \aslash)}
\def\jbloa{{\tr}_{+}(\jslash\, \bslash\, \lslash_1 \, \aslash)}
\def\ibPb{{\tr}_{+}(\islash\, \bslash\, \Pslash \, \bslash)}
\def\ijltb{{\tr}_{+}(\islash\, \jslash\, \lslash_2 \, \bslash)}

\def\ijloa{{\tr}_{+}(\islash\, \jslash\,  \lslash_1 \, \aslash)}
\def\ijblt{{\tr}_{+}(\islash\, \jslash\,  \bslash \, \lslash_2)}

\def\jakb{{\tr}_{+}(\jslash\, \aslash\, \kslash \, \bslash)}
\def\iakb{{\tr}_{+}(\islash\, \aslash\, \kslash \, \bslash)}

\def\tofo{{\tr}_{+}(\onslash\, \thslash\, \twslash \, \foslash)}
\def\foto{{\tr}_{+}(\onslash\, \thslash\, \foslash \, \twslash)}
\def\tofi{{\tr}_{+}(\onslash\, \thslash\, \twslash \, \fislash)}
\def\fito{{\tr}_{+}(\onslash\, \thslash\, \fislash \, \twslash)}

\def\lrangle#1#2{\langle #1\,#2\rangle}

\def\Li{{$\rm Li}_2$}
\def\eps{\epsilon}
\def\epsuv{{\epsilon_{\rm \mbox{\tiny UV}}}}
\let\bm=\bibitem
\let\la=\label

\def\npb#1#2#3{Nucl. Phys. {\bf{B#1}} #3 (#2)}
\def\plb#1#2#3{Phys. Lett. {\bf{#1B}} #3 (#2)}
\def\prl#1#2#3{Phys. Rev. Lett. {\bf{#1}} #3 (#2)}
\def\prd#1#2#3{Phys. Rev. {D \bf{#1}} #3 (#2)}
\def\cmp#1#2#3{Comm. Math. Phys. {\bf{#1}} #3 (#2)}
\def\cqg#1#2#3{Class. Quantum Grav. {\bf{#1}} #3 (#2)}
\def\nppsa#1#2#3{Nucl. Phys. B (Proc. Suppl.) {\bf{#1A}}#3 (#2)}
\def\ap#1#2#3{Ann. of Phys. {\bf{#1}} #3 (#2)}
\def\ijmp#1#2#3{Int. J. Mod. Phys. {\bf{A#1}} #3 (#2)}
\def\rmp#1#2#3{Rev. Mod. Phys. {\bf{#1}} #3 (#2)}
\def\mpla#1#2#3{Mod. Phys. Lett. {\bf A#1} #3 (#2)}
\def\jhep#1#2#3{J. High Energy Phys. {\bf #1} #3 (#2)}
\def\atmp#1#2#3{Adv. Theor. Math. Phys. {\bf #1} #3 (#2)}
\newcommand{\EQ}[1]{\begin{equation} #1 \end{equation}}
\newcommand{\AL}[1]{\begin{subequations}\begin{align} #1 \end{align}\end{subequations}}
\newcommand{\SP}[1]{\begin{equation}\begin{split} #1 \end{split}\end{equation}}
\newcommand{\ALAT}[2]{\begin{subequations}\begin{alignat}{#1} #2 \end{alignat}
                        \end{subequations}}
\def\beqa{\begin{eqnarray}}
\def\eeqa{\end{eqnarray}}
\def\beq{\begin{equation}}
\def\eeq{\end{equation}}
\def\sst{\scriptscriptstyle}
\def\thetabar{\bar\theta}
\def\Tr{{\rm Tr}}
\def\one{\mbox{1 \kern-.59em {\rm l}}}
 \def\Nh{\hat{N}}

\newcommand{\half}{{\textstyle {1 \over 2}}}

\def\a{\alpha}      \def\da{{\dot\alpha}}
\def\b{\beta}       \def\db{{\dot\beta}}
\def\c{\gamma}  \def\G{\Gamma}  \def\cdt{\dot\gamma}
\def\d{\delta}  \def\D{\Delta}  \def\ddt{\dot\delta}
\def\e{\epsilon}        \def\vare{\varepsilon}
\def\f{\phi}    \def\F{\Phi}    \def\vvf{\f}
\def\h{\eta}
\def\k{\kappa}
\def\l{\lambda} \def\L{\Lambda}
\def\m{\mu} \def\n{\nu}
\def\o{\omega}
\def\p{\pi} \def\P{\Pi}
\def\r{\rho}
\def\s{\sigma}  \def\S{\Sigma}
\def\t{\tau}
\def\th{\theta} \def\Th{\Theta} \def\vth{\vartheta}
\def\X{\Xeta}
\def\z{\zeta}
\def\de{\partial}

\def\cA{{\cal A}} \def\cB{{\cal B}} \def\cC{{\cal C}}
\def\cD{{\cal D}} \def\cE{{\cal E}} \def\cF{{\cal F}}
\def\cG{{\cal G}} \def\cH{{\cal H}} \def\cI{{\cal I}}
\def\cJ{{\cal J}} \def\cK{{\cal K}} \def\cL{{\cal L}}
\def\cM{{\cal M}} \def\cN{{\cal N}} \def\cO{{\cal O}}
\def\cP{{\cal P}} \def\cQ{{\cal Q}} \def\cR{{\cal R}}
\def\cS{{\cal S}} \def\cT{{\cal T}} \def\cU{{\cal U}}
\def\cV{{\cal V}} \def\cW{{\cal W}} \def\cX{{\cal X}}
\def\cY{{\cal Y}} \def\cZ{{\cal Z}}

\def\ua{\underline{\alpha}}
\def\ub{\underline{\phantom{\alpha}}\!\!\!\beta}
\def\uc{\underline{\phantom{\alpha}}\!\!\!\gamma}
\def\um{\underline{\mu}}
\def\ud{\underline\delta}
\def\ue{\underline\epsilon}
\def\una{\underline a}\def\unA{\underline A}
\def\unb{\underline b}\def\unB{\underline B}
\def\unc{\underline c}\def\unC{\underline C}
\def\und{\underline d}\def\unD{\underline D}
\def\une{\underline e}\def\unE{\underline E}
\def\unf{\underline{\phantom{e}}\!\!\!\! f}\def\unF{\underline F}
\def\unm{\underline m}\def\unM{\underline M}
\def\unn{\underline n}\def\unN{\underline N}
\def\unp{\underline{\phantom{a}}\!\!\! p}\def\unP{\underline P}
\def\unq{\underline{\phantom{a}}\!\!\! q}
\def\unQ{\underline{\phantom{A}}\!\!\!\! Q}
\def\unH{\underline{H}}

\def\As {{A \hspace{-6.4pt} \slash}\;}
\def\bs {{b \hspace{-6.4pt} \slash}\;}
\def\Ds {{D \hspace{-6.4pt} \slash}\;}
\def\ds {{\del \hspace{-6.4pt} \slash}\;}
\def\ss {{\s \hspace{-6.4pt} \slash}\;}
\def\ks {{ k \hspace{-6.4pt} \slash}\;}
\def\ps {{p \hspace{-6.4pt} \slash}\;}
\def\pas {{{p_1} \hspace{-6.4pt} \slash}\;}
\def\pbs {{{p_2} \hspace{-6.4pt} \slash}\;}
\def\Ps {{P \hspace{-6.4pt} \slash}\;}
\def\Qs {{Q \hspace{-6.4pt} \slash}\;}

\def\Fh{\hat{F}}
\def\Vh{\hat{V}}
\def\Xh{\hat{X}}
\def\ah{\hat{a}}
\def\xh{\hat{x}}
\def\yh{\hat{y}}
\def\ph{\hat{p}}
\def\xih{\hat{\xi}}
\def\psit{\tilde{\psi}}
\def\Psit{\tilde{\Psi}}
\def\tht{\tilde{\th}}
\def\lt{\tilde{\lambda}}
\def\hl{\hat{\lambda}}
\def\hlt{\hat{\tilde{\lambda}}}
\def\llt{\tilde{l}}
\def\At{\tilde{A}}
\def\Qt{\tilde{Q}}
\def\Rt{\tilde{R}}
\def\Nt{\tilde{N}}

\def\at{\tilde{a}}
\def\st{\tilde{s}}
\def\ft{\tilde{f}}
\def\pt{\tilde{p}}
\def\qt{\tilde{q}}
\def\vt{\tilde{v}}
\def\nt{\tilde{n}}

\def\delb{\bar{\partial}}
\def\bz{\bar{z}}
\def\bD{\bar{D}}
\def\bB{\bar{B}}

\def\bk{{\bf k}}
\def\bl{{\bf l}}
\def\bp{{\bf p}}
\def\bq{{\bf q}}
\def\br{{\bf r}}
\def\bx{{\bf x}}
\def\by{{\bf y}}
\def\bR{{\bf R}}
\def\bV{{\bf V}}

\def\d{\delta}\def\D{\Delta}\def\ddt{\dot\delta}
\def\pa{\partial} \def\del{\partial}
\def\xx{\times}
\def\uno{\mbox{1 \kern-.59em {\rm l}}}
\def\trp{^{\top}}
\def\inv{^{-1}}
\def\dag{{^{\dagger}}}
\def\pr{^{\prime}}
\def\lan{\langle}
\def\ran{\rangle}
\def\rar{\rightarrow}
\def\lar{\leftarrow}
\def\lrar{\leftrightarrow}
\newcommand{\0}{\,\!}      
\def\one{1\!\!1\,\,}
\def\im{\imath}
\def\jm{\jmath}
\newcommand{\tr}{\mbox{tr}}
\newcommand{\slsh}[1]{/ \!\!\!\! #1}
\def\vac{|0\rangle}
\def\lvac{\langle 0|}
\def\hlf{\frac{1}{2}}
\def\ove#1{\frac{1}{#1}}
\def\Box{\square}
\def\ZZ{\mathbb{Z}}
\def\CC#1{({\bf #1})}
\def\bcomment#1{}
\def\bfhat#1{{\bf \hat{#1}}}
\def\VEV#1{\left\langle #1\right\rangle}
\newcommand{\ex}[1]{{\rm e}^{#1}} \def\ii{{\rm i}}
\def\rr{{\rm r}} \def\rs{{\rm s}}\def\rv{{\rm v}}
\def\ri{{\rm i}}\def\rj{{\rm j}}
\newcommand{\lrbrk}[1]{\left(#1\right)}
\newcommand{\sfrac}[2]{{\textstyle\frac{#1}{#2}}}

\def\Li{{\rm Li}_2}

\font\mybb=msbm10 at 12pt
\def\bb#1{\hbox{\mybb#1}}

\font\myBB=msbm10 at 18pt
\def\BB#1{\hbox{\myBB#1}}

\begin{document}

\begin{flushright}
QMUL-PH-12-10 \\
WIS/05/12-MAY-DPPA
\end{flushright}

\vspace{8pt}

\begin{center}

{\Large \bf A note on  amplitudes in   $\mathcal{N}=6$ superconformal }
\\
\vspace{0.4cm}
{\Large \bf    Chern-Simons  theory  }

\vspace{16pt}

{\mbox {\bf  Andreas Brandhuber$^{a, b}$,  Gabriele Travaglini$^{a}$ and   Congkao Wen$^{a}$}}%
\footnote{
{\tt  \{ \tt \!\!\!a.brandhuber, g.travaglini, c.wen\}@qmul.ac.uk}
}


\begin{quote}
{\small \em
\begin{itemize}
\item[\ \ \ \ \ \ $^a$]
\begin{flushleft}
Centre for Research in String Theory\\
School of Physics and Astronomy\\
Queen Mary University of London\\
Mile End Road, London E1 4NS, United Kingdom
\end{flushleft}
\item[\ \ \ \ \ \ $^b$]
Department of Particle Physics and Astrophysics\\
Weizmann Institute of Science,
Rehovot 76100, Israel
\end{itemize}
}
\end{quote}


\vspace{60pt} {\bf Abstract}
\end{center}

\noindent
We establish a connection between tree-level superamplitudes in ABJM theory and leading singularities  
associated to special three-particle cuts of one-loop superamplitudes where one of the tree amplitudes entering the cut is a four-point amplitude. Using these relations, we show that certain intriguing similarities between one-loop  and tree-level superamplitudes observed recently become completely manifest. 
This connection is reminiscent of a similar relation in the maximally supersymmetric gauge theory in four dimensions, where the sum of two-mass hard and one-mass box coefficients of a one-loop amplitude equals the corresponding tree-level amplitude. As an application, we present a very simple re-derivation of the six-point superamplitude and  calculate the eight-point  superamplitude at one loop in  ABJM theory.

\setcounter{page}{0}
\thispagestyle{empty}
\newpage


\setcounter{tocdepth}{4}
\hrule height 0.75pt
\tableofcontents
\vspace{0.8cm}
\hrule height 0.75pt
\vspace{1cm}

\setcounter{tocdepth}{2}


\setcounter{footnote}{0}

 \section{Introduction}

The construction and study of three-dimensional superconformal field theories has seen a major surge in recent years. 
There are several motivations for this interest, two of the main ones being the search for a description of the low-energy physics of membranes in M-theory (M2-branes), 
and the importance of finding new examples of the AdS/CFT correspondence. 
Superconformal Chern-Simons theories (SCS) were first explored in \cite{schwarz} as possible
candidates of theories of M2-branes, and this idea was brought to fruition in \cite{bl}
and \cite{gustavsson} 
where the first construction of a three-dimensional theory with maximal, ${\mathcal N}=8$ superconformal symmetry (BLG) was presented.
Subsequently, in \cite{abjm} a large class of ${\mathcal N}=6$ superconformal theories
based on $U(N)_{+k} \times U(N)_{-k}$ Chern-Simons matter%
\footnote{The subscripts denote the levels of the Chern-Simon terms which come with opposite signs for the two gauge group factors.} 
theories (ABJM) was constructed. These theories were conjectured to be dual to type IIA string theory on $AdS_4 \times \bb{C}\bb{P}^3$ or M-theory on $AdS_4 \times S^7/\bb{Z}_k$, and to be the low-energy worldvolume theories of M2-branes near 
$\bb{Z}_k$ orbifolds.

Given the large number of new examples of three-dimensional theories with holographic duals, it is important 
to understand both sides of the duality, explore and exploit similarities with the dualities of four-dimensional theories, 
in particular $\mathcal{N}=4$ supersymmetric Yang-Mills (SYM), and more importantly identify any novel features. 
What one is particularly interested in are any hidden structures or symmetries, such as dual conformal symmetry or integrability, that are not necessarily manifest 
at the Lagrangian level and often manifest themselves only at the level of physical quantities like correlation functions, 
scattering amplitudes or Wilson loops.

In the context of the ${\rm AdS}_4/{\rm CFT}_3$ correspondence, integrability of the classical sigma model was studied in
\cite{ads4int}, while at weak coupling the dilatation operator of gauge-invariant operators
could be identified with a spin chain Hamiltonian in \cite{minahan}.
A corresponding Bethe equation for the all-loop anomalous dimensions was put forward in 
\cite{gromovvieira} (see also \cite{bethe}) and the Bethe ansatz proposal has been tested in many examples. 
Interestingly, it was found in \cite{gromovvieira, mcloughlin} that for particular operators,
namely twist operators, the quantum corrections in ABJM theory and $\mathcal{N}=4$ SYM are  related by a simple map of the 't Hooft couplings $\lambda_{\mathcal{N}=4} \to \left[ h(\lambda_{\mathrm{ABJM}}) \right]^2$ with 
$h(\lambda) = \lambda + c_1 \lambda^3 + c_2 \lambda^5 + \cdots$ at weak coupling \cite{gaiotto}, 
and $\lambda_{\mathrm{ABJM}}=N/k$. This observation is of particular interest as the anomalous dimension of twist-two operators  controls the leading infrared (IR) singularities of scattering amplitudes via its relation to the cusp anomalous dimension, and might hint at a deeper connection between the amplitudes of  these theories.

In this paper we will focus on aspects of the S-matrix of the ABJM theory, and we now briefly summarise some of the known results which will be relevant for our work. 
To begin with, various tree amplitudes up to eight points have been constructed explicitly 
using Feynman diagrams, on-shell recursion relations or Grassmannian formulations
\cite{Agarwal:2008pu, Bargheer:2010hn, Huang:2010qy, Gang:2010gy}. At one loop, 
the four-point amplitude was shown to vanish in  \cite{Agarwal:2008pu,Chen:2011vv},
while higher-point amplitudes are non-vanishing at the same loop order as was shown recently in 
\cite{yutin, Bargheer:2012cp} using triple cuts, and in \cite{Bianchi:2012cq} from supergraphs. 
At two loops, only the four-point amplitude has been  computed so far \cite{Chen:2011vv, Bianchi:2011dg}, 
with a result which surprisingly  matches the one-loop amplitude in $\mathcal{N}=4$ SYM. 
This equality  was later extended  to include all orders in the expansion in the dimensional regularisation parameter $\epsilon$ in  \cite{Bianchi:2011aa}. 
Wilson loops and a  potential duality  to amplitudes similar  to that in $\cN=4 $ SYM \cite{am,dks,bht}
were explored in 
\cite{Henn:2010ps,  Bianchi:2011rn, Wiegandt:2011uu, Bianchi:2011dg}. 
The result that lightlike polygonal Wilson loops at one loop are in fact zero  \cite{Henn:2010ps,  Bianchi:2011rn}
hints at some important issues to be understood in constructing such a duality, echoed by the appearance of singularities in formulating a fermionic T-duality \cite{ilyaetal}. 
On the other hand, it was found in \cite{Wiegandt:2011uu} that the two-loop result for the $2n$-gon Wilson loops
matches numerically the expression of  the corresponding one-loop result in $\mathcal{N}=4$ SYM \cite{bht} -- 
an intriguing  connection to the maximally supersymmetric theory in four dimensions. 

There are other properties familiar from  $\cN=4$ SYM that also appear in ABJM theory, 
notably dual superconformal symmetry \cite{dhks} and Yangian symmetry \cite{dhp}. 
These have been studied in  \cite{Bargheer:2010hn, Huang:2010qy, Gang:2010gy,lee}, 
and more recently the subtle breaking of some of these symmetries was understood in  
\cite{Bargheer:2012cp}. 

Novel features  of amplitudes in ABJM (compared to $\mathcal{N}=4$ SYM) are that amplitudes always have an even number of external states, and one-loop amplitudes are IR finite, in addition to being UV-finite. Intuitively, one can argue that this softer IR behaviour is related to the fact that the gluon in Chern-Simons theory is non-dynamical, and hence cannot appear on external lines, and the gluon propagator goes as $\sim 1/p$. More concretely, one-loop amplitudes can be expanded in terms of one-loop triangle functions \cite{Bargheer:2012cp}, a fact that is expected because of dual conformal invariance. In dimensional regularisation these triangles vanish if at least one of the external momenta is on-shell and otherwise give a finite result. But there is a further, more physical argument why IR divergences should be absent. 
General theorems guarantee that IR divergences cancel when IR-safe quantities such as cross sections are considered. In particular the virtual IR divergence associated with the interference of a one-loop $2n$-point amplitude with an $2n$-point tree amplitude has to be cancelled by the real emission IR divergence associated with (the modulus squared
of) the $(2n+1)$-point tree amplitude integrated over the phase space of an extra soft/collinear particle. 
The absence of $(2n+1)$-point tree amplitudes in ABJM then implies the IR finiteness of all one-loop amplitudes.%
\footnote{Strictly speaking these statements depend on the use of dimensional regularisation. When a different regulator is used, e.g.~adding a Yang-Mills term to the action, the gluon can become dynamical and then even one-loop amplitudes can become IR divergent as the regulated theory will contain amplitudes with an odd number of external legs. It would be interesting to analyse this issue in more detail.}
However a two-loop $2n$-point amplitude can be combined with (the square of) a $(2n+2)$-point tree amplitude, 
implying that two-loop amplitudes in ABJM are IR divergent, as indeed found in \cite{Chen:2011vv, Bianchi:2011dg}.


A related surprising feature of one-loop amplitudes in ABJM is  their remarkable similarity to their tree-level counterpart \cite{yutin, Bargheer:2012cp, Bianchi:2012cq}. This feature can be understood as a consequence of Yangian invariance  \cite{Agarwal:2008pu}
and its violation \cite{Bargheer:2012cp, Bianchi:2012cq} in the six-point case, but we feel it would be useful to have a more direct connection between trees and loops that may allow to understand if a similar pattern continues for eight and more external particles. 
This is the main question to which we provide an answer in this paper. 

\begin{figure}[h]
\scalebox{1.0}{
\centerline{\includegraphics[height=4.5cm]{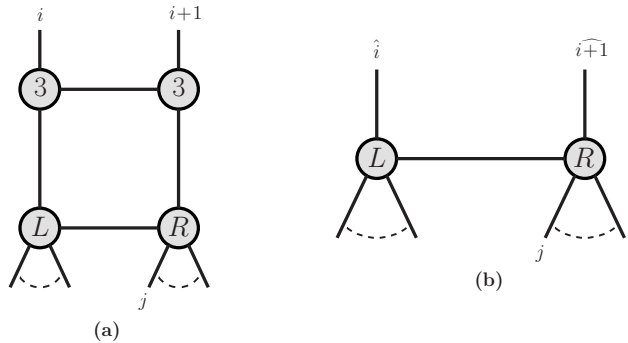} }
}
\caption{\it
Two-mass hard quadruple cut and the corresponding recursive diagram in $\cN=4$ SYM.  }
 \label{fig1}
 \end{figure}

To achieve this goal, we will  draw inspiration from $\cN=4$ SYM, where a remarkable connection exists between tree amplitudes and certain sums of coefficients of one-loop amplitudes when expanded in a basis of box functions. 
Specifically, it was observed in \cite{dissolving} that 
the sum of coefficients of all so-called two-mass hard box functions, i.e.~boxes which have two adjacent massless corners with momenta 
$p_i$ and $p_{i+1}$, as in Figure \ref{fig1}(a), 
equals twice the tree-level amplitude in $\cN=4$ SYM. This is true for any choice of the two adjacent massless legs, thus there are $n$ such equations for an $n$-point amplitude. 
These relations were later proved in \cite{simplest}, where it was shown that they arise as particular combinations of the IR consistency equations. The reader might at this point object that due to the absence of IR divergences at one loop in ABJM theory, this strategy is bound to fail. However, there is an alternative  proof of the relations of \cite{dissolving} which was found in \cite{bht2}, relying on the one-loop dual conformal anomaly of amplitudes, whose expression  was conjectured in \cite{dhks5} and proved in \cite{bht3}. Specifically, it was found in  \cite{bht2} that in order to satisfy the anomalous dual conformal Ward identities, one-loop supercoefficients must obey certain linear equations, $n$ of which are precisely the relations between tree amplitudes and one-loop two-mass hard and one-mass box coefficients described in 
\cite{dissolving}. 

At this point we mention a key ingredient in our story, namely the connection between  quadruple cuts involving two adjacent massless corners -- from which one obtains the coefficients that feature in the relations of \cite{dissolving}, 
and tree-level recursion relations  \cite{bcf, bcfw}.  In the context of $\cN=4$ SYM, this relation was first noticed in \cite{bcf}, and in \cite{simplest} it was shown how one can map each quadruple cut of this type directly  to a BCFW recursive diagram. In particular 
the two cut legs depicted as vertical lines in Figure \ref{fig1}(a) morph into BCFW shifted legs, see  Figure \ref{fig1}(b). 
In this way, the relations of \cite{dissolving} are proved because the sum of two-mass hard coefficients is converted into the sum of BCFW recursive diagrams which in turn is known to give the tree amplitude.
%
%
%
%
%
%
%
%

Motivated by this, we will consider special triple cuts in three-dimensional ABJM theory  
where one of the  participating amplitudes is a four-point tree-level superamplitude, while the remaining two can have any (even) number of particles.  
Note that in ABJM the four-point amplitude is the $\it smallest$ amplitude and plays a similar fundamental role as the three-point amplitude in $\cN=4$ SYM. In \cite{Bargheer:2012cp} the special role of four-point tree amplitudes was emphasised and their superconformal anomalies, which are localised on collinear configurations, 
were studied in detail and shown to be the source of superconformal anomalies of six-point tree and one-loop amplitudes.%
\footnote{Incidentally,  we wish to point out the similar role played by four-point tree amplitudes in unitarity cuts of one-loop superamplitudes 
in $\cN=4$ SYM. Indeed it is the particular class of two-particle cuts including a four-point tree amplitude that is related to IR divergences and responsible for the one-loop dual conformal anomaly  as pointed out in \cite{bht3}.}
There are two solutions to the triple-cut equations, and hence two contributions to the supercoefficients, paralleling the 
two solutions $z_{1,2}$ for the $z$ variable defining the shifts in the three-dimensional recursion  relation \cite{Gang:2010gy}. 
In $\cN=4$ SYM, the sum of the two contributions from the two quadruple cut solutions gives 
the manifestly PT-invariant form of the supersymmetric BCFW recursion relation. It contains two terms which happen to be equal as  a consequence of the large-$z$ behaviour of the superamplitude \cite{simplest}, leading to the factor of two alluded to earlier. 
 
What happens in the three-dimensional  ABJM theory? We find something rather surprising here.  
If combined with a positive sign, these two ``sharpened" leading singularities, in the language of \cite{sharpen}, 
give the coefficient of a triangle function. 
However, it is their difference which reproduces a  BCFW recursive diagram for the corresponding tree-level amplitude. 
Turning things around, one can combine the two terms resulting from the  evaluation of the same recursive diagram at the two residues $z_{1,2}$ with plus or minus sign, in one case leading to the tree-level amplitude, in the other to a supercoefficient. 
The fact that these leading singularities can be combined in different ways leads one to suspect that they should be separately Yangian invariant. In \cite{Gang:2010gy}  dual conformal symmetry was shown to hold diagram by diagram in the recursion relation, following a strategy similar to that of  \cite{bhtdcs}. 
In fact, following the proof of  \cite{Gang:2010gy}  one can easily convince oneself that each leading singularity is separately dual conformal covariant.  At six points, the triple cuts at one loop contain only four-point amplitudes  \cite{Bargheer:2012cp}. It is then clear that the one-loop six-point amplitude must be closely related to tree amplitudes, because of the connection between the special triple cuts where one of the cut-amplitudes is a four-point amplitude and recursive diagrams that we have outlined above. 
It is also clear that the same connection can be used to relate any leading singularity where one of the participating amplitudes is a four-point amplitude to recursive diagrams, and in particular  to fully determine the eight- and ten-point superamplitudes at one loop. 
We demonstrate the efficiency of this strategy by
calculating  the one-loop eight-point superamplitude explicitly.


The rest of the paper is organised as follows.  In Section 2 we review some  basic facts about amplitudes in ABJM, their
description in on-shell superspace and summarise their factorisation properties at tree level. In Section 3 we present the
main observation of our paper, namely that particular triple cuts 
where at least one of the three tree-amplitudes appearing in the cut is a four-point amplitude
are in one-to-one correspondence with recursive diagrams for tree amplitudes. As a byproduct, we write down 
a more compact form of the tree-level recursion relations. 
In Section 4 we re-derive the one-loop six-point amplitude and in Section 5 we show how our connection to recursive diagrams can also be used to derive  
the  eight-point  one-loop superamplitude.


\section{Lightning review of superamplitudes in ABJM}

Here we briefly review some facts about ABJM theory and, in particular,  its scattering amplitudes. 
The field content consists of four complex scalar fields 
$\phi^A$, and four complex fermions $\psi^\alpha_A$, where $\a = 1,2$ is a spin index and $A=1, \ldots , 4$ is an $SU(4)$ $R$-symmetry group index, with the scalars (fermions) transforming in its (anti)fundamental representation. 
Furthermore,  
$(\phi^A, \psi_A^\alpha)$ transform in the $(N, \bar{N})$ representation of the $U(N) \times U(N)$ gauge group, while the complex conjugate fields $(\bar\phi_A, \bar\psi^A_\alpha)$ transform in the $(\bar{N}, N)$. Finally, the gauge fields $A_\mu$ and $\hat{A}_\mu$ are described by a Chern-Simons action and hence they have  no on-shell degrees of freedom,  i.e.~gauge fields do not appear in the external states.

The matter content of ABJM theory can be efficiently described using  $\cN=3 $ superspace  \cite{Bargheer:2010hn}. In this set-up, one introduces  a Nair superfield 
$\Phi( \l, \eta)$ to describe particles and its Grassmann Fourier transform $\bar\Phi( \l, \eta) $  for the antiparticles: 
\beq
\label{phi}
\Phi( \l, \eta) \ = \ \phi^4 (\l)  \, + \, \eta^A \psi_A (\l) \, + \, {1\over 2} \eps_{ABC} \eta^A \eta^B \phi^C (\l) \, + \,  {1\over 3!} \eps_{ABC} \eta^A \eta^B \eta^C \psi_4 (\l) \ , 
\eeq
\beq
\label{phibar}
\bar\Phi( \l, \eta) \ = \ \bar\psi^4 (\l) \, + \, \eta^A \bar\phi_A (\l) \, + \, {1\over 2} \eps_{ABC} \eta^A \eta^B \bar\psi^C (\l) \, + \,  {1\over 3!} \eps_{ABC} \eta^A \eta^B \eta^C \bar\phi_4 (\l) \ . 
\eeq
Here $\eta^A$, $A=1,2,3$ are Grassmann variables transforming in the fundamental representation of $SU(3)$. Note that $\Phi$ is bosonic, whereas $\bar\Phi$ is fermionic. Momenta of on-shell particles are written in  the three-dimensional spinor helicity formalism as products of commuting spinors $\l^\a$ as 
\beq
p^{\a \b}  \ :=  \ \l^\a \l^\b\, . 
\eeq
For the reader's convenience, we summarise in Appendix A useful facts about the spinor helicity formalism, along with our conventions.  
In terms of the $(\l, \eta)$  variables,  supersymmetry generators have a very simple form \cite{Bargheer:2010hn}, for instance
\beq
Q^{\alpha A} \ = \ \l^\a \eta^A \, , \qquad Q^\a_A \ = \ \l^\a {\partial \over \partial \eta^A} \ , 
\eeq
while the $R$-symmetry generators act as 
\beq
R^{AB} \ = \ \eta^A \eta^B \, , \qquad R^A_B \ = \ \eta^A  {\partial \over \partial \eta^B} - {1\over 2} \delta^A_B \, , \qquad
R_{AB} \ = \  {\partial \over \partial \eta^A}  {\partial \over \partial \eta^B}  
\ . 
\eeq
Colour-ordered superamplitudes, which are the subject of this paper, are denoted by 
\beq
\cM \ = \cM (\bar\Phi_1, \Phi_2, \bar\Phi_3, \ldots, \bar\Phi_{n-1}, \Phi_n)
\ , 
\eeq
where particular states can be chosen by taking the appropriate power in the $\eta$ expansion according to \eqref{phi} or \eqref{phibar} depending on whether the corresponding field appearing in $\cM$ is unbarred (particle) or barred (antiparticle). Note that 
$n$ is even, a fact that follows simply from gauge invariance once one recalls that the superfields $\Phi$ and $\bar\Phi$ carry colour indices 
$\Phi^a_{\, \bar{b}}$ and $\Phi^{\bar{a}}_{\, b}$, where (un)barred indices are associated to the first  (second) $U(N)$ group. 
In the following we will often use a simplified notation $\cM (\bar{1}, 2, \bar{3}, \ldots , \overline{n-1}, n)$ to denote superamplitudes. 

An important difference with the four-dimensional helicity formalism  is that the little group of the Lorentz group is now discrete. Indeed, particles' momenta $p^{\a\b} = \ \l^\a \l^\b$ and supermomenta  $ q^{\a A} = \l^\a \eta^A$,  
are invariant under 
\beq
\l \to - \l \, , \qquad \eta \to - \eta
\, . 
\eeq
Note that under this transformation one has $\Phi \to \Phi$, while $\bar\Phi \to -   \bar\Phi$, and hence 
\beq
\label{little}
\cM ( \cdots ; - \l_i,  - \eta_i;  \cdots ) \ = \ (-)^i \, \cM ( \cdots;  \l_i  , \eta_i;  \cdots ) 
\ , 
\eeq
i.e.~the superamplitude flips sign or not according to whether $i$ labels an  antiparticle or a particle. 

Finally, we observe that an $n$-point amplitude has  fermionic degree $3 n /2$,   in sharp contradistinction with the standard formulation of four-dimensional $\cN=4$ SYM using Nair's chiral superspace, where the fermionic degree is related to the MHV degree of the superamplitude. Perhaps the only amplitude reminiscent of the four-dimensional MHV superamplitude is the four-point amplitude, whose expression is \cite{Agarwal:2008pu}
\beq
\label{fourtree}
\cM_4 (\bar{1}, 2, \bar{3},4) \ = \  {\delta^{(3)} ( \sum_{i=1}^4 \l_i \l_i)\, \delta^{(6)} ( \sum_{i=1}^4 \l_i \eta_i) \over \lan 12 \ran \lan 23 \ran 
 }
\ . 
\eeq
Note the presence of a $\delta^{(6)} ( \sum_{i=1}^4 \l_i \eta_i)$ of supermomentum conservation, which makes supersymmetry manifest. 

We will round up this section with a brief summary of the factorisation properties of amplitudes in ABJM theory. We will mainly focus on tree amplitudes in three types of kinematic limits:
$(a)$ soft limit $p_i \to 0$,  $(b)$ collinear limit $p_i \sim c \, p_{i+1}$ and its generalisations, and $(c)$ multi-particle factorisation $P^2_{ij} \to 0$ where $P_{ij} := p_i + \cdots + p_j$. 
Let us begin with the collinear limit of the four-particle amplitude \eqref{fourtree} and take e.g.~momenta 1 and 2 collinear, 
i.e.~$\langle 1 2\rangle \to 0$. Momentum conservation and on-shell conditions however force momenta 3 and 4 also to be collinear with 1 and 2, and hence all spinor brackets vanish. From this we conclude that the four-point superamplitude vanishes as $\langle 1 2 \rangle$ in this limit. Similarly one can show that \eqref{fourtree} vanishes in the soft limit.
There is however one subtlety, associated with peculiar anti-collinear configurations such as
$p_1+p_2=0$ and $p_3+p_4=0$ \cite{Bargheer:2010hn}. In this  case $\lambda_2 = i \lambda_1$ and $\lambda_4=i \lambda_3$, but $\lambda_1$ and $\lambda_3$ are unrelated. Thus, not all spinor brackets appearing in the expansion of the fermionic delta-function in \eqref{fourtree} vanish, 
and the four-point amplitude diverges as $\sim \langle 1 2\rangle^{-1}$. This divergence is associated with a zero-momentum gluon exchange diagram, as explained in \cite{Agarwal:2008pu,Bargheer:2010hn, Bargheer:2012cp}.

Multiparticle factorisation only occurs for six-point and higher-point amplitudes. In a multi-particle  limit, an amplitude factorises 
into a product of two tree amplitudes times a propagator in the channel that goes on-shell. Since  only amplitudes with an even number of external legs are non-zero,  we have the following pattern \cite{Bargheer:2010hn}:  
\begin{enumerate} 
\item[{\bf A.}]
Channels with an odd number of momenta: 
$P^2 = (p_1 + \ldots + p_{1+2 k})^2 \to 0$: in this  case the limit of the amplitude is singular and factorisation occurs  as
\beq
\hspace{-1cm}\mathcal{M}_{2n}(1, \ldots, n) \to \int\!\!d^3 \eta_P \ \mathcal{M}_{2 k }(\{1,\ldots, 1+2k\},P)\, \frac{1}{P^2} \,\mathcal{M}_{2n-2k}(-P,\{2+2k,\ldots, n\})
\, + \, \mathrm{finite}
\eeq 
\item[{\bf B.}] Channels with an even number of momenta: 
$P^2 = (p_i + \ldots + p_{i + 2k + 1})^2 \to 0$, which includes collinear limits for $k=0$.  In this case no singularity arises, as this would require factorisation onto two tree amplitudes with an odd number of external legs. 
\end{enumerate}
It is interesting to note that  in ABJM the usual collinear singularities of amplitudes  are absent, a fact which is closely related to the absence of one-loop IR divergences. Similarly, a soft limit with otherwise generic kinematics does not lead to a singularity.
One can also consider multi-collinear limits such as a triple collinear limit, which is a special case of  the case A above with $k=1$. The four-point amplitude appearing in the factorisation limit vanishes as $\langle 1 2 \rangle$ in this case  (we assume here that $i=1$), however this gets multiplied with the singular propagator $\sim \langle 1 2 \rangle^{-2}$ which leads in total to a pole of the form $\langle 1 2 \rangle^{-1} \times \mathcal{M}_{2n-2}$.


\section{Connection between anomalous three-particle cuts and recursive diagrams}

One-loop amplitudes in superconformal Chern-Simons theory are believed to be finite. This is a consequence of the conjectured dual conformal symmetry property of the theory \cite{Gang:2010gy}
-- see the Introduction for alternative explanations. Indeed, because of the symmetry  only  scalar triangle functions will appear in the expansion of a one-loop amplitude. One-mass and two-mass triangles in dimensional regularisation are zero when evaluated in  $D=3$, and we are left with three-mass triangles, which are dual conformal invariant if multiplied with an appropriate normalisation factor.  However, lacking enough independent region momenta, the only way this normalised  integral function can be dual conformal invariant is by being a constant 
(see Appendix B for details). 

In the following we will write general one-loop amplitudes as a linear combination of unnormalised, scalar triangle functions,  
\beq
\label{3mtrrr}
\cM \ = \ \sum_{K_1, K_2, K_3} \mathcal{C}_{K_1 K_2 K_3} \, \mathcal{I}(K_1,  K_2 , K_3)
\ , 
\eeq
where $\mathcal{C}_{K_1 K_2 K_3} $ are supercoefficients and $ \mathcal{I}(K_1,  K_2 , K_3)$ denote the three-mass triangles integrals 
\beqa
\label{3mts}
\cI^{\rm 3m}(K_1, K_2, K_3) &:= &  \int\!\!{d^3 l}  \, {1 \over( l^2 + i \varepsilon)  ( (l +K_1)^2 + i \varepsilon) ( (l +K_1 + K_2)^2 + i \varepsilon)  }
\nonumber \\
&=& { -i \,  \pi^3 \over  \, \sqrt{- (K_1^2+ i \varepsilon)}   \sqrt{ -(K_2^2+ i \varepsilon)}  \sqrt{-( K_3^2 + i \varepsilon)} }
\  ,
\eeqa
with $K_i^2 \neq 0$ for all $i$. 
The expression for these triangles was obtained in \cite{kaza,gra,dav},  and we provide an independent derivation 
using Mellin Barnes representations in Appendix B.

\subsection{Supercoefficients from triple cuts}

The supercoefficients can be calculated by applying generalised unitarity 
\cite{Bern:1997sc,Britto:2004nc}  or, specifically in three-dimensions, three-particle cuts. This  strategy was pursued in \cite{Chen:2011vv} where the four-point amplitude at one and two loops were calculated, and very recently in \cite{yutin, Bargheer:2012cp}, where three-particle cuts were performed in order to calculate the six-point amplitude at one loop. This amplitude was concurrently derived in \cite{Bianchi:2012cq} using supergraphs. 
%
%
%
%
\begin{figure}[h]
\scalebox{1}{
\centerline{\includegraphics[height=4.5cm]{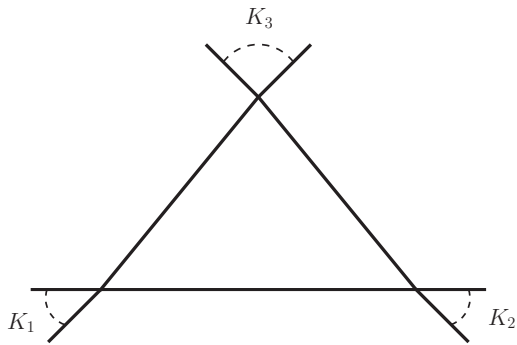} }
}
\caption{\it
A generic three-mass triangle function. In three dimensions, one-mass and two-mass triangles vanish. }
 \label{trianglefun}
 \end{figure}
%
%
%
%

From triple cuts, one can derive an equation  which determines  $ \cC_{K_1 K_2 K_3}$, 
\beq
 \cC_{K_1 K_2 K_3}\, \cdot J   \ = \ \sum_{l^\ast}  
\big|{\rm det}  \left( { \partial f_i(l_a) \over \partial l_{a\mu} } \right)^{-1}_{l=l^\ast} \big|  \, \int\!\!d^3\eta_a\, d^3\eta_b\, d^3\eta_c\ \cM_1 \cM_2 \cM_3
\  , 
\eeq
where $l^\ast$ denotes the two solutions to the equations 
\beq 
\label{eq:cuts}
 f_1(l_a) = l_a^2=0, \quad f_2(l_a) = (l_a +K_1)^2=0, \quad f_3(l_a)= (l_a+ K_1 +K_2)^2 =0 
\ , 
\eeq
 $K_i$, $i = 1,2,3$,  are the external momenta at each corner of the triangle, see Figure \ref{trianglefun},
and  $J$ is the Jacobian
\beq
J \ :=  \ 
\sum_{l^\ast} \big|  {\rm det}  \left( { \partial f_i(l_a) \over \partial l_{a\mu} } \right)^{-1}_{l=l^\ast}\big|
\  .
\eeq
$l^\ast$  can  expressed as  \cite{Bargheer:2012cp}
\beq
l^\ast \ = \ \alpha K_1 + \beta K_2 + \gamma K_{\perp}
\ , 
\eeq
with  $ K_{\perp}^\mu := \eps^{\mu \nu \rho} K_{1 \nu} K_{2 \rho}$. One also finds that 
$\gamma = s   \sqrt{- K_1^2 K_2^2 K_3^2} / (2 K_{\perp}^2)$ where $s= \pm$ for the two solutions.%
\footnote{ 
The expressions for $\alpha$ and $\beta$ can be found in  \cite{Bargheer:2012cp} and will be immaterial in the following. Note that in our conventions  $\eta^{\mu \nu}= (+ -  - )$. }
The determinant is easily evaluated to be
\beqa
\big|{\rm det}  \left( { \partial f_i(l_a) \over \partial l_{a\mu} } \right)_{l=l^\ast} \big| 
&= &
%
%
8\,\big| {\rm det} (l^\ast_\mu, K_{1 \mu}, K_{2 \mu}) \big|\ = \ 
8\,  \big| \gamma (\epsilon_{\mu\nu\rho} K_1^\nu K_2^\rho)^2\big|
\nonumber \\ \cr
&= &4  \, \sqrt{-K_1^2 K_2^2 K_3^2}
\ , 
\eeqa
 and, hence, performing the sum over the two solutions,  we get
\beq
J \ = \ {1\over 2} {1 \over \sqrt{- K_1^2 K_2^2 K_3^2 } }
\ . 
\eeq
We conclude that 
\beq
\label{supercoeff}
\cC_{K_1 K_2 K_3} \ = \ {1\over 2} \, \sum_{l^\ast}   \,  \int\!\!d^3\eta_a\, d^3\eta_b\, d^3\eta_c\ \cM_1 \cM_2 \cM_3
\  .
\eeq
%
%
%
\begin{figure}[h]
\centerline{\includegraphics[height=6cm]{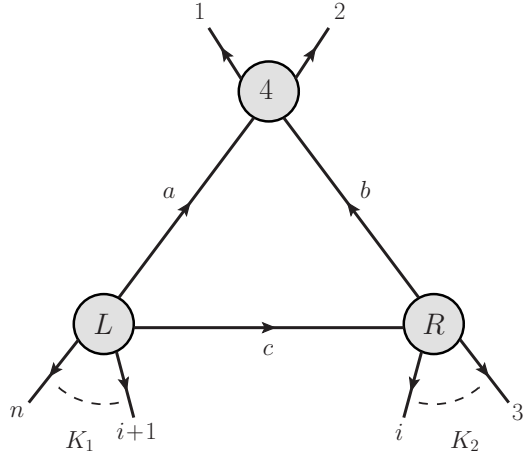} } 
\caption{\it
The particular three-particle cut considered in Section \ref{section-anom} which we use to evaluate the supercoefficient $\cC_{12; i}$. Note that $K_1 := p_{i+1} + \cdots + p_n$, and $K_2 := p_3 + \cdots + p_i$.  }
 \label{figcut}
 \end{figure}
%
%
%

\subsection{Solution for the anomalous triple-cut equations} 
\label{section-anom}

In the following  we will focus on a particular class of supercoefficients, where one of the three massive corners (denoted earlier as $K_3$) 
contains only two particles, i.e.~the corresponding amplitude at that corner is a four-point amplitude. 
We  call these triple cuts ``anomalous" because of the presence of a four-point tree amplitude, which is the source of anomalies of the Yangian generators  \cite{Bargheer:2012cp}. 
Momentum conservation reads now 
\beq
K_1 + K_2 + P_{12}  \ = \ 0 
\ , 
\eeq
with $P_{12} := p_1 + p_2$. 
The corresponding three-particle cut is depicted in Figure \ref{figcut}, where the cut momenta are $l_a$, $l_b$ and $l_c$, with $l_a^2 = l_b^2 = l_c^2=0$. We also have $l_a + l_b = P_{12}$. 

In order to establish a connection between this particular class of triple cuts and recursion diagrams we  will now solve the triple cut conditions in a way that closely parallels the on-shell conditions for the shifted legs in the recursion relation. 
To this end, we note that the conditions $l_a^2 = l_b^2 = 0$ (in the notation of Figure \ref{figcut}) can be satisfied by setting $(l_a)_{\a \b}  := \hat{\l}_{a; \a}\hat{\l}_{a; \b}$, $(l_b)_{\a \b}  := \hat{\l}_{b; \a}\hat{\l}_{b; \b}$ with 
\beqa
\hat{\l}_a &= &x \l_1 - y \l_2 \ , 
\nonumber \\
\hat{\l}_b &=& y \l_1 + x \l_2 \ . 
\eeqa
Momentum conservation implies $l_a + l_b = (x^2 + y^2) (\l_1 \l_1 + \l_2 \l_2) $, and hence $x^2 + y^2 = 1$. 
We will solve this condition by setting \cite{Gang:2010gy} 
\beqa
x & = & {1\over 2} ( z + z^{-1}) \ , 
\nonumber \\
y & = & {1\over 2 i } ( z - z^{-1}) \ , 
\eeqa
or 
\beq
\label{shiftl}
\begin{pmatrix} 
\hat\l_a \\ \hat\l_b 
\end{pmatrix} \ = \ 
R(z) \, \begin{pmatrix} 
\l_1 \\ \l_2 
\end{pmatrix}
\ , 
\eeq
where 
\beq
R(z) \ = \ 
\begin{pmatrix} 
{1\over 2} (z + z^{-1})  & - {1\over 2i }  (z - z^{-1}) 
\\ \cr
{1\over 2i }  (z - z^{-1})  & 
{1\over 2}  (z + z^{-1}) 
\end{pmatrix} 
\ . 
\eeq
Note that $R^T \, R = \uno$. Finally, $z$ can be determined by solving the remaining on-shell condition 
\beq
\label{remai}
l_c^2 =(l_a + K_1)^2 = 0
\ . 
\eeq
This turns out to be a biquadratic equation in $z$, as can be seen in the following way \cite{Gang:2010gy}. One notices that 
\beq
\hat\lambda_a \ = \ {1\over 2} \Big[ z ( \l_1 + i \l_2) + z^{-1} ( \l_1 - i \l_2) \Big]
\ , 
\eeq
and introducing 
\beq
q^{\a \b} \ := \ {1\over 4} (\l_1 + i \l_2)^\a (\l_1 + i \l_2)^\b \ , 
\qquad 
\tilde{q}^{\a \b} \ := \ {1\over 4} (\l_1 - i \l_2)^\a (\l_1 - i \l_2)^\b \ , 
\eeq
one can rewrite 
\beq
l_a \ := \ \hat{\l}_a \hat{\l}_a  \ = \ z^2 q \, +\,  z^{-2} \tilde{q} \, + \, {1\over 2} ( p_1 + p_2) 
 \ . 
 \eeq
Eqn.~\eqref{remai} then takes the form, 
\beq
\label{biquadratica}
a z^{-2} + b + c z^2 = 0 
\ , 
\eeq
with
\beq
\label{coeff}
a \ = \ 2 (\tilde{q} \cdot K_1) \ , \qquad b \ = \ - K_1 \cdot K_2 \ , \qquad c \ = \ 2 (q \cdot K_1) \ , 
\eeq
whose solutions can be cast in the form%
\footnote{After using the identity $(q\cdot K_1) (\tilde{q}\cdot K_1) = (1/16) \big[ (K_1 \cdot K_2 )^2 - K_1^2 K_2^2 \big]$.}
 \cite{Gang:2010gy}
\beq
\label{explsol}
z_1^2 \ = \ {  K_1 \cdot K_2 + \sqrt{ K_1^2 K_2^2} \over 4  (q \cdot K_1) } \,  , \qquad 
z_2^2 \ = \ {  K_1 \cdot K_2 - \sqrt{ K_1^2 K_2^2} \over 4  (q \cdot K_1) }
\ .
\eeq
There is an interesting way to rewrite \eqref{explsol} making use of a peculiar representation  for  generic non-null three-dimensional vectors $K$, discussed in \eqref{nonnull} of Appendix A.  
We define    
\beq
\label{K1K2}
K_{1\, ab}\  := \ \xi_{(a} \mu_{b)}\, , \qquad  K_{2\, ab} \ :=\  \xi_{(a}^\prime \mu_{b)}^\prime\, , 
\eeq
and use  \eqref{squared}  to write  
 $ K_1^2 K_2^2 =(1/16) \lan \xi \mu \ran^2 \lan \xi^\prime \mu^\prime \ran^2$.
The solutions $z_{1,2}$ can therefore be re-cast in the following form,%
\footnote{The precise definition of $z_1^2$ and $z_2^2$ in \eqref{explsol2} is chosen in order to agree with the explicit solutions in  the six-point case 
given in \eqref{six-ptsol}.}
\beq
\label{explsol2}
z_1^2 \ = \ {  \lan \xi \mu^\prime \ran \lan \mu \xi^\prime \ran  \over \lan \l_1 + i \l_2 | \, K_1\, | \l_1 + i \l_2 \ran  }
 \,  , \qquad 
z_2^2 \ = \ { \lan \xi \xi^\prime \ran  \lan \mu \mu^\prime \ran \over \lan \l_1 + i \l_2 | \, K_1\, | \l_1 + i \l_2 \ran }
\ , 
\eeq
where $z_{1,2}$  correspond to the two choices  $K_1 \cdot K_2  \mp (1/4) \lan \xi \mu \ran \lan \xi^\prime \mu^\prime \ran$, respectively.
Note that, effectively, \eqref {explsol2} can be obtained from \eqref{explsol} by performing the replacement 
\beq
\sqrt{ K_1^2 K_2^2} \to - (1/4) \lan \xi \mu \ran \lan \xi^\prime \mu^\prime \ran
\ . 
\eeq
Very interestingly,   all square roots have disappeared in this new representation of the solution given in \eqref{explsol2}.
This feature will be useful in the following.

\subsection{Anomalous triple cuts and the associated recursive diagrams} 

\begin{figure}[h]
\centerline{\includegraphics[height=6cm]{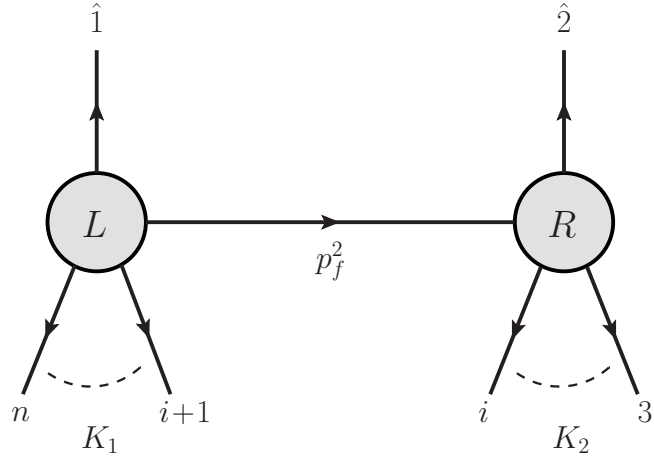} } 
\caption{\it
The recursive diagram which we associate with the three-particle cut in Figure \ref{figbcfw}. We also set $K_1:= p_{i+1} + \cdots + p_n $, and $K_2 := p_3 + \cdots + p_i$.}
 \label{figbcfw}
 \end{figure}

At this point let us compare the triple-cut equations we have just considered, which we momentarily  rewrite as 
\beq
\label{eq1}
l_c^2 \ = \ 0 \ , \qquad (l_c + K_1)^2 \ = \ 0 \ , \qquad (l_c - K_2)^2 \ = \ 0 \
\ , 
\eeq
to the equations determining the BCFW shift in the recursion diagram in Figure \ref{figbcfw}. The latter are
$\hat{p}_f^2 = \hat{p}_1^2 = \hat{p}_2^2  =0$, or 
\beq
\label{eq2}
\hat{p}_f^2 \ = \ 0 \ , \qquad (\hat{p}_f + K_1)^2 \ = \ 0 \ , \qquad (\hat{p}_f - K_2)^2 \ = \ 0 \
\ .
\eeq
Note that \eqref{eq1} and  \eqref{eq2} are identical in form, and so will the solutions for $l_c$ and $\hat{p}_f$. This is the first evidence of an underlying  connection between the triple-cut diagram in Figure \ref{figcut} and the recursive diagram in Figure \ref{figbcfw}.

Next, we move on to evaluating explicitly the triple cut in Figure \ref{figcut}. We denote  the product of the three amplitudes (before evaluating them on the solution of the cut equations) with the slightly simplified notation  $\mathcal{C}_{12; i}(z)$, where the label $i$ is introduced   in Figure \ref{figcut}. Using \eqref{supercoeff}, the supercoefficient is then given by 
\beq
\label{supercoeffreloaded}
 \cC_{12; i}   \ =  \ {1\over 2}  \sum_{l^\ast}\mathcal{C}_{12; i}(l^\ast) \ = \ 
  {1\over 4}  \sum_{z\in\{\pm z_1, \pm z_2\}}\mathcal{C}_{12; i}(z) \ .
  \eeq 
 Note the appearance of an  extra factor of $1/2$ on the right-hand side of \eqref{supercoeffreloaded} compared to 
 \eqref{supercoeff}, which is due to the sum over equal and opposite roots of $z_i$, $i=1,2$.

The expression for $\mathcal{C}_{12; i}(z)$ is 
\beq
\mathcal{C}_{12; i}(z) \ = \ \int\!\!d^3\eta_a d^3\eta_b d^3\eta_c \,   \cM_4 (\bar{1}, 2, - \bar{b}, -a) \, 
\cM_R ( \bar{3}, \ldots, i,  - \bar{c}, b)
\cM_L (\overline{i+1} ,  \ldots, n, \bar{a}, c )  \, , 
\eeq   
where the four-point superamplitude is given in \eqref{fourtree}.  
Here we adopt  the short-hand notations $\cM (1, \ldots , n) = \cM (\l_1, \eta_1; \ldots ; \l_n, \eta_n)$ and  
$\cM( \cdots ; - a;  \cdots ) =  \cM(\cdots;   i \l_a, i \eta_a ; \cdots )$, where $-a$ indicates that the momentum and supermomentum of particle $a$ are $-\l_a\l_a$ and $-\l_a\eta_a$. 

In order to perform the Grassmann integrations efficiently we use the identity 
\beq
\delta^{(6)} \Big( \sum_{i=1}^2 \l_i \eta_i - \hat{\l}_a \eta_a - \hat{\l}_b \eta_b \Big)  \ = \ \lan 1 2 \ran^3 \delta^{(3)} ( \eta_a -\hat{\eta}_a )
\delta^{(3)} ( \eta_b -\hat{\eta}_b )
\ , 
\eeq
where 
\beq
\label{shifte}
\begin{pmatrix} 
\hat{\eta}_a \\ \hat{\eta}_b 
\end{pmatrix} \ = \ 
R(z) \, \begin{pmatrix} 
\eta_1 \\ \eta_2 
\end{pmatrix}
\ , 
\eeq
and we have also used $\lan \hat{a} \, \hat{b} \ran = \lan 1\, 2 \ran$. 
The integration over $\eta_a$ and $\eta_b$ is then performed trivially, and the particular triple cut we are considering becomes  
\beq
\mathcal{C}_{12; i}(z)\ = \ {\lan 1\, 2 \ran^2 \over i \lan 2  \, \hat{b}\ran}
\int\!d^3 \eta_c \ \cM_R(  \bar{3}  \ldots , i ,  - \bar{c} ,\hat{b}  ) \, \cM_L( \overline{i+1}, \ldots , n , \bar{\hat{a}}, c  ) 
\ , 
\eeq
where 
\beq
\lan 2\,  \hat{b} \ran = - {1\over 2i } \lan 1\, 2 \ran ( z - z^{-1})
\ . 
\eeq
Thus we arrive at the result 
\beq
\label{sopra}
\mathcal{C}_{12; i}  (z) \ = \ - 2\, {  \lan 12 \ran \over z - z^{-1}}   
\int\!d^3 \eta_c \ \cM_R(  \bar{3}  \ldots , i, - \bar{c} , \hat{b}  )  \, \cM_L( \overline{i+1}, \ldots , n , \bar{\hat{a}}, c ) 
\ . 
\eeq
The next step consists in performing the sum over the four solutions $(z_1, -z_1,  z_2, -z_2)$  to  \eqref{biquadratica} in order to obtain \eqref{supercoeffreloaded}.
To this end, we first note the property
 \cite{Gang:2010gy} 
\beq
\label{propertyg}
\cM_R(-z) \cM_L(-z) \ = \ - \cM_R(z) \cM_L(z)
\ , 
\eeq
which follows from $\hat{\l}_a$ and  $\hat{\l}_b$ being odd in $z$ and from \eqref{little}, 
together with the fact that particles 1 and 2 are adjacent in colour ordering.  
Using \eqref{propertyg}, we can rewrite the sum on the right-hand side of \eqref{supercoeffreloaded} as 
\beqa
\label{sopra2}
\sum_{z\in\{ \pm z_1, \pm z_2\}} \mathcal{C}_{12; i}  (z) &=& - 4\, {  \lan 12 \ran \over z_1 - z_1^{-1}}   
\int\!d^3 \eta_c \ \Big[\, \cM_R( \bar{3}  \ldots , i, - \bar{c} , \hat{b}    ) \, \cM_L(\overline{i+1}, \ldots , n , \bar{\hat{a}}, c   )  \Big]_{z=z_1} 
\nonumber \\
&+&       z_1 \rightarrow z_2\ . 
\eeqa   
%
%
%
%
%
%
At this point we would like to establish a connection between \eqref{sopra2} and a particular diagram in the BCFW recursion relation formulated  in \cite{Gang:2010gy}. 
More specifically, we now compare \eqref{sopra2} with the recursive diagram depicted in Figure \ref{figbcfw}, whose expression is given by 
 \cite{Gang:2010gy}  
\beqa
\label{recdiagr}
\mathcal{R}_{12; i} &=&  
 \int\!d^3 \eta_c \   {H(z_1, z_2) \over p_f^2}  \Big[ \cM_R(\bar{3}  \ldots , i, - \bar{c} , \hat{2}    )   \, \cM_L(\overline{i+1}, \ldots , n , \bar{\hat{1}}, c    )  \Big]_{z = z_1} \ +  \ 
(z_1 \leftrightarrow z_2)   
\nonumber \\
&:= & Y_{12; i}^{(1)} \ +\ Y_{12; i}^{(2)}
\ , 
\eeqa
where $p_f = p_2 + \cdots + p_i$ is the momentum in the internal propagator,  $(z_1, -z_1,  z_2, -z_2)$ are the four solutions to \eqref{eq2} (or equivalently  \eqref{biquadratica}), and 
\beq
H(z_1, z_2) := {z_1 ( z_2^2 - 1 ) \over z_1^2 - z_2^2}
\ . 
\eeq
Here we have also introduced shifted superspace variables  $\hat{\l}_{1,2}$ and $\hat\eta_{1,2}$ which are defined by formulae that are identical in form to \eqref{shiftl} and  \eqref{shifte}, namely
\beq
\label{shiftle}
\begin{pmatrix} 
\hat\l_1 \\ \hat\l_2 
\end{pmatrix} \ = \ 
R(z) \, \begin{pmatrix} 
\l_1 \\ \l_2 
\end{pmatrix}
\ , \qquad  
\begin{pmatrix} 
\hat{\eta}_1 \\ \hat{\eta}_2
\end{pmatrix} \ = \ 
R(z) \, \begin{pmatrix} 
\eta_1 \\ \eta_2 
\end{pmatrix}
\ .
\eeq
In other words, the solution $\hat{a} := ( \hat{\lambda}_a, \hat\eta_a)$ and $\hat{b} := ( \hat{\lambda}_b, \hat\eta_b)$ to the cut conditions  for the legs $a$ and $b$ in the the triple-cut diagram are precisely the same as the three-dimensional BCFW shifts. Note that a very similar relation between certain quadruple cuts with two adjacent three-particle amplitudes as depicted in Figure \ref{fig1} and on-shell recursive diagrams was observed in four dimensions already in \cite{bcf}. It is very interesting that the very same connection between maximal cuts and recursion relations emerges also in the three-dimensional theory considered here.
Furthermore,  from the viewpoint of the triple cuts the  non-linear BCFW shifts in three dimensions appear very naturally.

In order to simplify the notation, we define 
\beq
G(z) \ = \ {\lan 1\, 2\ran \over z - z^{-1}}
\ , 
\eeq 
and look for a relation between $H(z_1, z_2)$, and $G(z_1)$ and $G(z_2)$. 
After a short calculation making use of the explicit expressions for the solutions $z_{1,2}$,  we find  that%
\footnote{
Specifically, from \eqref{biquadratica} and \eqref{coeff} it follows  that  $ c(1 - z_1^2) (1-z_2^2) = p_f^2$ and $c(z_1^2 - z_2^2) = \sqrt{K_1^2 K_2^2}$. 
In the notation of \eqref{explsol2},
the latter equation reads $c(z_1^2 - z_2^2) = - (1/4) \lan \xi \mu\ran \lan \xi^\prime \mu^\prime \ran$.  }
\beq
\label{HG}
\frac{1}{p_f^2}{H(z_1, z_2)\over G(z_1)}  \ = \ - \frac{1}{p_f^2}{H(z_2, z_1)\over G(z_2)} \ = \ {1 \over \lan 1\, 2\ran  \sqrt{K_1^2 \, K_2^2}}
\ . 
\eeq
The minus sign in the first equality in \eqref{HG} is  very important -- if it were not present, we would ultimately be led to conclude that one-loop amplitudes are proportional to tree-level amplitudes, which is not the case.

We conclude by stating the precise relation between the recursive diagram and the triple-cut diagram considered so far, namely 
\beq
\label{A}
\mathcal{R}_{12; i} \ = \ -
{1 \over  4\,  \lan 1\, 2\ran  \sqrt{K_1^2 \, K_2^2} }   
\Big[\sum_{z= \pm z_1} \mathcal{C}_{12; i}  (z)  - \sum_{z= \pm z_2} \mathcal{C}_{12; i}  (z) \Big]
\ . 
\eeq
The sum $\sum_i \mathcal{R}_{12; i}$ over all possible recursive diagrams gives the tree-level superamplitude $\cM_{\rm tree}$, and hence we have derived the new relation 
\beq
\label{minus}
\cM_{\rm tree}\ = \ - \sum_i  {\sum_{z = \pm z_1} \mathcal{C}_{12; i}  (z) \, - \,  \sum_{z = \pm z_2} \mathcal{C}_{12; i}  (z)   \over4\,  \lan 1\, 2\ran  \sqrt{K_1^2 \, K_2^2} } 
\ , 
\eeq
where the sum is extended to all supercoefficients corresponding to the triple cuts in Figure \ref{figcut}, where one of the corners has fixed legs, say 1 and 2. 
This is one of the main results of this section. Using the notation \eqref{explsol2} we can re-write \eqref{A} and \eqref{minus} in a neater form manifestly free of square roots: 
\beq
\label{Abis}
\mathcal{R}_{12; i} \ = \
{ 1\over \lan 1\, 2\ran  \lan \xi \mu\ran \lan \xi^\prime \mu^\prime \ran  }   
\Big[\sum_{z =  \pm z_1} \mathcal{C}_{12; i}  (z)  - \sum_{z =  \pm z_2} \mathcal{C}_{12; i}  (z) \Big]
\ , 
\eeq
\beq
\label{minusBIS}
\cM_{\rm tree}\ = \ \sum_i   {\sum_{z = \pm z_1} \mathcal{C}_{12; i}  (z) \, - \,  \sum_{z =  \pm z_2} \mathcal{C}_{12; i}  (z)   \over\,  \lan 1\, 2\ran  \lan \xi \mu\ran \lan \xi^\prime \mu^\prime \ran } 
\ , 
\eeq
%
 Incidentally, we note that the BCFW recursion relation  can also be re-expressed in the interesting form%
\footnote{Recall that  $K_1 := p_{i+1} + \cdots + p_n$ and $K_2 = p_3 + \cdots + p_i$.} 
\beqa
\cM_{\rm tree} & = &  
\sum_i {1\over \sqrt{ K_1^2 K_2^2}}\,  \int\!\!d^3 \eta_c \   {1 \over z_1 - z_1^{-1}  }  \Big[\cM_R( \bar{3}  \ldots , i, - \bar{c} , \hat{2}    ) \,   \cM_L(\overline{i+1}, \ldots , n , \bar{\hat{1}}, c    )  \Big]_{z = z_1} 
\nonumber \\ \cr
&-&
(z_1 \rightarrow z_2)   
\ .
\eeqa  
%
Conversely, we can re-express the supercoefficient as a function of the two recursion diagram terms $Y_{12; i}^{(n)}$, $n = 1,2$ defined in  \eqref{recdiagr}.  Since 
\beq
G(z_1) \ = \ \lan12\ran \sqrt{K_1^2 K_2^2} \  {H(z_1, z_2) \over  p_f^2}\ , \qquad 
G(z_2) \ = \ - \lan12\ran \sqrt{K_1^2 K_2^2} \  {H(z_2, z_1) \over  p_f^2}\ ,
\eeq
we find, using \eqref{supercoeffreloaded}, 
\beq
\cC_{12; i}   \ =  
- \lan12\ran \sqrt{K_1^2 K_2^2} \ \Big( Y_{12; i}^{(1)} - Y_{12; i}^{(2)} \Big)
 \ , 
  \eeq 
  or
  \beq
  \label{supxm}
  \cC_{12; i}  \ =  
{1\over 4} \lan12\ran \lan \xi \mu\ran \lan \xi^\prime \mu^\prime \ran \ \Big( Y_{12; i}^{(1)} - Y_{12; i}^{(2)} \Big)
 \ . 
  \eeq
Multiplying this by the corresponding three-mass triangle \eqref{3mts} we obtain a contribution equal to  
\beq
\label{contri}
 \cC_{12; i} \, \cI_{12, K_1, K_2} \ = \ 
-i{  \pi^3\over 4}  { \lan12\ran \over \sqrt{- (P_{12}^2 + i \varepsilon)}}
{\lan \xi \mu \ran  \over  \sqrt{- (K_1^2 + i \varepsilon)}} { \lan \xi^\prime \mu^\prime \ran  \over   \sqrt{- (K_2^2 + i \varepsilon)}}
\ \Big( Y_{12; i}^{(1)} - Y_{12; i}^{(2)} \Big)
\ , 
\eeq
where we remind the reader that $K_1$ and $K_2$ are written in \eqref{K1K2} using three-dimensional spinor notation.

Eqns.~\eqref{supxm} and \eqref{contri} are the other main results of this section; these are practical formulae of immediate applicability in deriving higher-point one-loop superamplitudes, as we will illustrate in several examples in the next sections.

We conclude this section with two short comments. 

{\bf 1.}
 We have seen that the combinations $ Y_{12; i}^{(1)} \pm Y_{12; i}^{(2)}$ correspond to either recursion diagrams or supercoefficients. It was demonstrated in \cite{Gang:2010gy} that each recursive diagram is separately dual conformal invariant. 
 As it was pointed out in the Introduction, one can see using the same proof  that in fact each residue 
 $ Y_{12; i}^{(1)}$ and $Y_{12; i}^{(2)}$ is separately dual conformal invariant. 

{\bf 2.} 
We note the appearance in \eqref{contri} of  peculiar ratios of Lorentz-invariant angle brackets to square roots of kinematic invariants.
These correspond to the  sign functions detected in  \cite{yutin, Bargheer:2012cp,Bianchi:2012cq}, and will be made more explicit in the six-point case discussed below, see \eqref{segno}.

\section{The one-loop six-point superamplitude}

We can use the results of the previous section, specifically \eqref{supxm} and \eqref{contri},  to re-derive the six-point one-loop superamplitude, recently obtained in \cite{Bargheer:2012cp,Bianchi:2012cq}. In this case the two possible three-particle cuts involve only four-point amplitudes.  For six-point kinematics,  \eqref{explsol2} takes the simple form
\beq
\label{six-ptsol} 
z^2_1 \ = \ 2\, { P_{34} \cdot P_{56} + \lan 3 \, 4\ran \lan 5 \, 6 \ran \over \lan \l_1 + i \l_2 | \, P_{34}\, | \l_1 + i \l_2 \ran}\, ,\qquad
 z^2_2 \ = \ 2\, { P_{34} \cdot P_{56} - \lan 3 \, 4\ran \lan 5 \, 6 \ran \over \lan \l_1 + i \l_2 | \, P_{34}\, | \l_1 + i \l_2 \ran} \, .
\eeq
Here we have used $K_1 := P_{34}$, $K_2 := P_{56}$ and $\lan \xi \mu \ran  = - 2 i\, \lan 3\,4 \ran$,  $\lan \xi^\prime \mu^\prime \ran  = - 2 i \,\lan 5\,6 \ran$, with $P_{34}^2 = \lan 3\,4\ran^2$, $P_{56}^2 = \lan 5\,6\ran^2$.

%
%
%
%
\begin{figure}[h]
\scalebox{1}{
\centerline{\includegraphics[height=4.5cm]{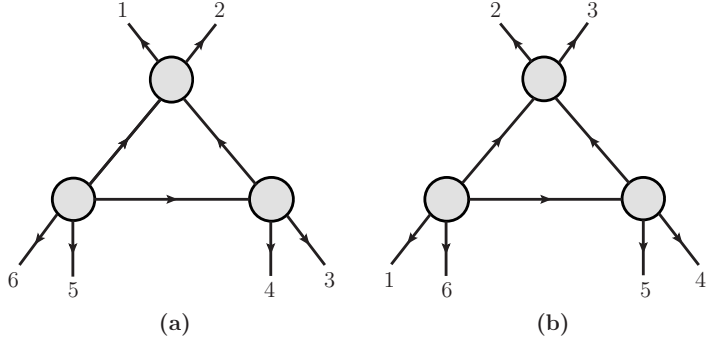} }
}
\caption{\it
The two contributions to the one-loop six-point amplitude. }
 \label{6-pt}
 \end{figure}
%
%
%
%

Using \eqref{six-ptsol}, it is straightforward to find the six-point tree-level amplitude from BCFW recursion relations%
\beqa \label{sixtree}
\cM_{\rm tree}(\bar{1}, 2, \bar{3}, 4, \bar{5}, 6 ) &:=& Y^{(1)}_{12;4} \ + \ Y^{(2)}_{12;4} 
\nonumber \\ 
\cr
 &=& {\delta^{(3)}(P) \delta^{(6)}(Q) \over P^2_{24}} \bigg[ { \delta^{(3)}(\epsilon_{ijk} \lan j \, k \ran \eta_i  - 
i \, \epsilon_{\bar{i} \bar{j} \bar{k}} \lan \bar{j} \, \bar{k} \ran \eta_{\bar{i}} ) 
 \over (\lan 2| P_{34}| 5 \ran + i \lan 3 \, 4\ran \lan 6 \, 1\ran)
 (\lan 1| P_{23}| 4 \ran + i \lan 2 \, 3\ran \lan 5 \, 6\ran) } 
 \nonumber \\
 \cr  
 &+ &  {\delta^{(3)}(\epsilon_{ijk} \lan j \, k \ran \eta_i  + 
i \, \epsilon_{\bar{i} \bar{j} \bar{k}} \lan \bar{j} \, \bar{k} \ran \eta_{\bar{i}} ) 
 \over (\lan 2| P_{34}| 5 \ran - i \lan 3 \, 4\ran \lan 6 \, 1\ran)
 (\lan 1| P_{23}| 4 \ran - i \lan 2 \, 3\ran \lan 5 \, 6\ran) }  \bigg] \, ,
\eeqa 
where
$i \, , j \, , k =  2 \, , 3 \, , 4 \, ,$ and $ \bar{i} \, , \bar{j} \, , \bar{k} =  5 \, , 6 \, , 1 \, .$

We can now write down the expression for  the corresponding  one-loop supercoefficient $\cC_{12;4}$ from the triple cut in 
Figure \ref{6-pt}(a) using \eqref{supxm},
\beqa
\cC_{12;4} &=& -\lan 1 \, 2  \ran  \lan 3 \, 4  \ran  \lan  5 \, 6\ran  \Big( Y^{(1)}_{12;4} - Y^{(2)}_{12;4} \Big) \, .
\eeqa
It was observed in \cite{Bargheer:2012cp} that the combination $Y^{(1)}_{12;4} - Y^{(2)}_{12;4}$ is in fact equal to the shifted tree-level amplitude $i\, \cM_{\rm tree} (\bar{6}, 1, \bar{2}, 3, \bar{4}, 5 )$.%
\footnote{This fact can be easily understood by comparing the BCFW diagram with the same shift for these two different amplitudes, $\cM_{\rm tree} (\bar{1}, 2, \bar{3}, 4, \bar{5}, 6 )$ and $\cM_{\rm tree} (\bar{6}, 1, \bar{2}, 3, \bar{4}, 5 )$. Similar but slightly more complicated relations may be obtained from BCFW for higher-point tree-level amplitudes. } 
Hence,  
\beqa
\label{c124} 
\cC_{12;4} &=& -i\, \lan 1 \, 2  \ran  \lan 3 \, 4  \ran  \lan  5 \, 6\ran  \cM_{\rm tree} (\bar{6}, 1, \bar{2}, 3, \bar{4}, 5 ) \, .
\eeqa
Similarly, we find for the supercoefficient $\cC_{23;5}$ of Figure \ref{6-pt}(b)
\beqa
\label{c235}
\cC_{23;5} &=& -i\, \lan 2 \, 3  \ran  \lan 4 \, 5  \ran  \lan  6 \, 1\ran  \cM_{\rm tree} (\bar{6}, 1, \bar{2}, 3, \bar{4}, 5 ) \, . 
\eeqa
Note that here the same tree-level amplitude $ \cM_{\rm tree} (\bar{6}, 1, \bar{2}, 3, \bar{4}, 5 )$ appears both in \eqref{c124} and \eqref{c235} -- 
in the two cases it appears as a  BCFW recursion relation for the same amplitude but with a different shift.

The final result for the one-loop superamplitude is obtained by  multiplying the supercoefficients by the corresponding three-mass triangle function  as in \eqref{contri}, with the result 
\beqa
\cM^{(1)}(\bar{1}, 2, \bar{3}, 4, \bar{5}, 6 ) &=& \cC_{12;4} \, \cI_{12, 34, 56} \, + \, 
 \cC_{23; 5}\, \cI_{23, 45, 61}  
\nonumber \\
\cr
&=&  i  \pi^3 \, \mathcal{S}(p) \,  \cM_{\rm tree} (\bar{6}, 1, \bar{2}, 3, \bar{4}, 5 ) \, ,
\eeqa
where the prefactor $\mathcal{S}(p)$ is
\beqa
\mathcal{S}(p) =  {\rm sgn}( \lan 1 \, 2 \ran ) {\rm sgn}( \lan 3 \, 4 \ran ) {\rm sgn}( \lan 5 \, 6 \ran ) + 
{\rm sgn}( \lan 2 \, 3 \ran ) {\rm sgn}( \lan 4 \, 5 \ran ) {\rm sgn}( \lan 6 \, 1 \ran ) \, , 
\eeqa
and  
\beq
\label{segno}
{\rm sgn} \big( \lan k \, l \ran  \big)\ := \  - i {\lan k \, l \ran \over \sqrt{ -(  \lan k \, l \ran^2+ i \varepsilon)}}
\ . 
\eeq
This is in agreement with the recent calculations of \cite{Bargheer:2012cp,Bianchi:2012cq}.

\section{Eight-point and ten-point superamplitudes at one loop from recursive diagrams}

There are two more superamplitudes whose triple cuts will always involve at least one four-point superamplitude, namely the eight-point and the ten-point superamplitudes. Using the procedure outlined in the previous sections, it is clear that such amplitudes can be expressed in terms of tree-level recursive diagrams evaluated on the different pole solutions $z_1 $ and $z_2$ for each case. As observed earlier, evaluating a recursive diagram on $z_i$ leads to a dual conformal invariant result even before summing over the two solutions $z_i^2$, $i=1,2$. Beyond ten points, there will be triple cuts involving three amplitudes with more than four legs; these will be genuinely new terms which have to be evaluated separately. 

In order to make this discussion more concrete, we illustrate it in the eight-point case. The ten-point superamplitude  can be addressed in the same way.

%
%
%
\begin{figure}[h]
\scalebox{1}{
\centerline{\includegraphics[height=4.5cm]{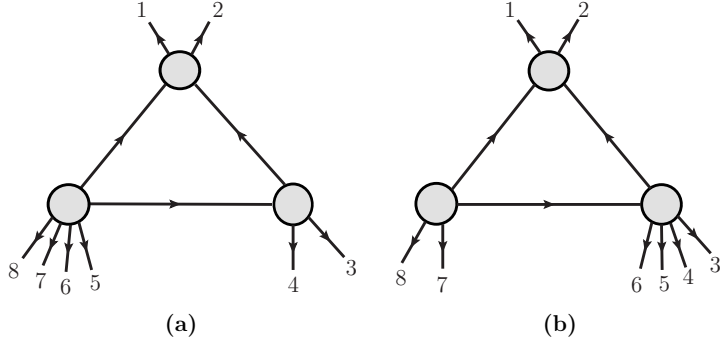} }
}
\caption{\it
Two of the eight triple-cut diagrams contributing to the eight-point superamplitude at one loop. The six remaining diagrams are obtained by cyclically shifting the particle labels by one, four, and five units. }
 \label{8-pt-cut}
 \end{figure}
%
%
%

%
%
%
\begin{figure}[h]
\scalebox{1}{
\centerline{\includegraphics[height=4.5cm]{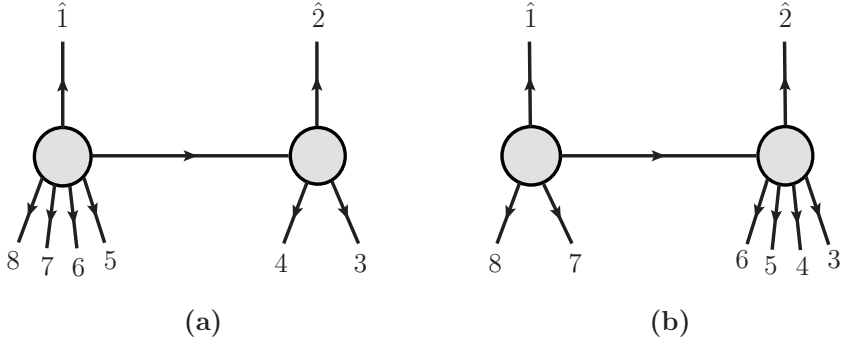} }
}
\caption{\it
The two recursive diagrams associated with the triple-cut diagrams shown in Figure \ref{8-pt-cut}. }
 \label{8-pt-rec}
 \end{figure}
%
%
%

At eight points, there are eight independent triple cuts to consider, two of which are depicted in Figure \ref{8-pt-cut}. The remaining six are obtained by shifting the labels by one, four, and five units.  According to the discussion of the previous sections, summarised by \eqref{supxm} and \eqref{contri},  
the two cut diagrams in Figure \ref{8-pt-cut} can be put in correspondence with the two recursive diagrams in Figure \ref{8-pt-rec}. 
In particular, the recursive diagram in Figure \ref{8-pt-rec}(a) has the expression%
\footnote{We evaluate explicitly these recursive diagrams and provide expressions for the functions $Y$ in Appendix C.} 
\beq
\cR^{(a)}_{12; 4} \ := \ Y^{(1)}_{12;4} \, + \, Y^{(2)}_{12;4}
\ , 
\eeq
and using  \eqref{contri}, the corresponding expression for the triple-cut diagram in Figure \ref{8-pt-cut}(a) multiplied by the corresponding triangle function will be 
\beq
\cC_{12; 4} \, \cI_{12, 34, 5678} \ = \ i \pi^3\,  \cS_{12;4} \Big( Y^{(1)}_{12;4} \, - \, Y^{(2)}_{12;4} \Big)
\ . 
\eeq
Here we denote by $Y^{(i)}_{12;4}$ the result of the evaluation of the recursive diagram in Figure \ref{8-pt-rec}(a) on the solution $z_i^2$, $i=1,2$, where $z_i^2$ are given in \eqref{explsol2}. The prefactor $ \cS_{12; 4} $ can be read off from \eqref{contri}, and has the form
\beq
\cS_{12;4} \ = \ -{  1\over 4}  { \lan12\ran \over \sqrt{- (P_{12}^2 + i \varepsilon)}}
{\lan \xi \mu \ran  \over  \sqrt{- (K_1^2 + i \varepsilon)}} { \lan \xi^\prime \mu^\prime \ran  \over   \sqrt{- (K_2^2 + i \varepsilon)}}
\ . 
\eeq
 In this case $K_1 = p_5 + \cdots + p_8 := P_{58}$, whereas $K_2 = p_3 + p_4$. Hence we can replace $ \lan \xi^\prime \mu^\prime \ran  = - 2i \lan 34 \ran$, and 
 \beq
 \cS_{12;4}  \ = \ {  i\over 2}  { \lan12\ran \over \sqrt{- (P_{12}^2 + i \varepsilon)}}
{\lan \xi \mu \ran  \over  \sqrt{- (P_{58}^2 + i \varepsilon)}} { \lan 34\ran  \over   \sqrt{- (P_{34}^2 + i \varepsilon)}}
\ . 
\eeq
 $\xi$ and $\mu$  are indirectly defined through $P_{58}^{ab} := \xi^{(a} \mu^{b)}$.  
 
 The contribution from Figure \ref{8-pt-cut}(b), multiplied by the appropriate triangle function, is 
 \beq
\cC_{12; 6} \, \cI_{12, 3456, 78} \ = \ i \pi^3\,  \cS_{12;6} \Big( Y^{(1)}_{12;6} \, - \, Y^{(2)}_{12;6} \Big)
\ , 
\eeq
where  the functions $Y^{(i)}_{12;6}$ correspond to the recursive diagrams in Figure  \ref{8-pt-rec}(b) evaluated on the two solutions $z_i^2$, $i=1,2$. The prefactor $\cS_{12;6}$ is now given by 
\beq
 \cS_{12;6}  \ = \ {  i\over 2}  { \lan12\ran \over \sqrt{- (P_{12}^2 + i \varepsilon)}}
{\lan 78 \ran  \over  \sqrt{- (P_{78}^2 + i \varepsilon)}}
{\lan \xi^\prime \mu^\prime \ran  \over  \sqrt{- (P_{36}^2 + i \varepsilon)}} \ , 
\eeq
where $P_{36}^{ab}  := \xi^{\prime (a} \mu^{\prime b)}$.

The final result is then obtained by summing eight contributions and reads
\beqa
\cM^{(1)}_8(\bar{1}, \ldots , 8)  &=&
 i \pi^3\, \Big[  \cS_{12;4} \Big( Y^{(1)}_{12;4} \, - \, Y^{(2)}_{12;4} \Big)
 \ + \   \cS_{12;6} \Big( Y^{(1)}_{12;6} \, - \, Y^{(2)}_{12;6} \Big)
\nonumber \\ \cr
&+&   \cS_{23;5} \Big( Y^{(1)}_{23;5} \, - \, Y^{(2)}_{23;5} \Big)
 \ + \    \cS_{23;7} \Big( Y^{(1)}_{23;7} \, - \, Y^{(2)}_{23;7} \Big)
 \nonumber \\ \cr
&+&  \cS_{56;8} \Big( Y^{(1)}_{56;8} \, - \, Y^{(2)}_{56;8} \Big)
 \ + \    \cS_{56;2} \Big( Y^{(1)}_{56;2} \, - \, Y^{(2)}_{56;2} \Big)
 \nonumber \\ \cr
&+&   \cS_{67;1} \Big( Y^{(1)}_{67;1} \, - \, Y^{(2)}_{67;1} \Big)
 \ + \   \cS_{67;3} \Big( Y^{(1)}_{67;3} \, - \, Y^{(2)}_{67;3} \Big)
\Big]
\ . 
\eeqa
Note the sum of certain pairs of $Y^{(1)}$ and $Y^{(2)}$ gives the eight-point tree-level amplitude, 
\beqa
\cM_{\rm tree}(\bar{1},2, \ldots , \bar{7}, 8) = Y^{(1)}_{ij;k} \, + \, Y^{(2)}_{ij;k} \, + \, Y^{(1)}_{ij;k\!+\!2} \, + \, Y^{(2)}_{ij;k\!+\!2} 
\ ,
\eeqa
namely they are the same amplitude with a different BCFW shift. We present an explicit calculation of tree-level eight-point amplitude in Appendix C.

The calculation of the ten-point amplitude would largely parallel the steps presented above
but we will not present the technical details here.
We only comment that in all 20 possible triple cuts at least one four-point amplitude appears and, hence, the
procedure goes through straightforwardly as for the six- and eight-point cases.
For twelve- and higher-point amplitudes in addition to the anomalous triple cuts also genuine non-anomalous triple cuts will appear which involve products of three amplitudes with more than four legs. 
We believe that the representation of the momenta $K_i$ appearing in the corners of the triangles using \eqref{nonnull} will be beneficial to derive compact expressions for completely general one-loop amplitudes%
\footnote{In particular note that also the solutions of the generic, non-anomalous triple cut (i.e.~when $K_3$ is the sum of an arbitrary number of momenta) can be represented in a form free of square roots. All steps of the derivation in Section \ref{section-anom} go through 
if we express $K_3$ as  $K_{3}^{ab} = \rho^{(a} \overline{\rho}^{b)}$ with $\rho$ complex,  and simply replace 
$\lambda_1 +i \lambda_2 \to  \rho$. In this way  we can still use the solution \eqref{explsol2}.}.



\section*{Acknowledgements}

We would like to thank Harald Ita, Tristan McLoughlin, Massimo Siani, Bill Spence and Gang Yang for very interesting discussions. 
This work was supported by the STFC Grant ST/J000469/1,  
``String theory, gauge
theory \& duality". AB thanks the ``Feinberg Foundation
Visiting Faculty Program" at the Weizmann Institute of Science and
Tel Aviv University for hospitality, and GT thanks the ETH Z\"{u}rich and the Niels Bohr International Academy for hospitality.  We would also like to thank the Newton Institute for Mathematical Sciences, Cambridge, for hospitality, the participants in the Newton Institute workshop {\it ``Recent Advances in Scattering Amplitudes"} for creating a stimulating atmosphere
and the speakers for giving wonderful talks. 
\appendix

\section{Conventions}

We present here a stenographic summary of our conventions. We work in signature $(+,-,-)$ and use the real Pauli matrices to relate momenta in vector and double-spinor notation,
\be
\sigma^0_{\a\b} = \left( \begin{matrix} 1  & 0 \\ 0 & 1 \end{matrix} \right) \ , \
\sigma^1_{\a\b} = \left( \begin{matrix} 0  & 1 \\ 1 & 0 \end{matrix} \right) \ , \
\sigma^2_{\a\b} = \left( \begin{matrix} 1  & 0 \\ 0 & -1 \end{matrix} \right) \ ,
\ee
such that a generic, possibly off-shell momentum can be written as
\be P_{\a\b}=p_\mu \sigma^\mu_{\a\b} = \left( \begin{matrix} E-p_y  & -p_x \\ -p_x & E+p_y \end{matrix} \right) \ .
\ee
Note that this is a symmetric matrix and hence any off-shell momentum can be written alternatively as the symmetrised
product of two two-spinors $\xi$,   $\mu$ as
\be
\label{nonnull}
P_{\a\b} = \xi_{(\a} \mu_{\b)} =\frac{1}{2} (\xi_\a \mu_\b+\xi_\b \mu_\a) \ ,
\ee
a useful fact that is used throughout the text.
Note that if we choose $\xi$ and $\mu$ to be real then there is a rescaling invariance in this
representation $\xi \to r \xi , \mu \to \mu/r$ with $r$ a non-zero real number. 
Alternatively we can choose the spinor variables to be complex, but then they are related by complex conjugation
$\mu = \overline{\xi}$ in order for the momenta to be real. This representation is invariant under $\xi \to e^{i \phi} \xi$,   $\mu \to e^{-i \phi} \mu$.

For on-shell momenta $p_\mu p^\mu = \det p_{\a\b}=0$, and  we simply set $\mu=\xi=\lambda$, which reduces the rank of the two-by-two matrix defined above and removes the rescaling invariance except for the reflection $\lambda \to -\lambda$. Therefore, we have
\be
p_{\a\b}\, =\, \lambda_\a \lambda_\b
\ , 
\ee
and we note that for positive energy  $\lambda$ must be real,  while for negative energy  it is purely imaginary.

Spinor variables can be contracted in an $SL(2,\bb{R})$ (Lorentz-)invariant fashion using the epsilon tensor $\epsilon_{\a\b}=\epsilon^{\a\b}$ with $\epsilon_{12}=+1$, which is also used to raise and lower spinor indices. The fundamental invariant of two spinor variables $\lambda$ and $\mu$  is defined as
\be
\langle \lambda \mu \rangle = \epsilon_{\a\b} \lambda^\a \mu^\b \ ,
\ee
in terms of which we can write any Lorentz-invariant momentum vector contractions, two very common examples being
\beqa
(p_1+p_2)^2 & = & \langle 1 2 \rangle^2 \ ,\nonumber \\
2 \epsilon^{\mu\nu\rho} p_{1\mu} p_{2\nu} p_{3\rho} & = & \tr (\sigma^\mu \sigma^\nu \sigma^\rho)p_{1\mu} p_{2\nu} p_{3\rho} = \langle 12 \rangle \langle 23 \rangle \langle 31 \rangle \ .
\eeqa
Here we have also introduced the short-hand notation $\langle \lambda_1 \lambda_2 \rangle \equiv \langle 12 \rangle$. 
Finally, we note that for a generic momentum written as in \eqref{nonnull}, we have 
\beq
\label{squared}
P^2  \ = \ -{1\over 4} \lan \xi \, \m \ran^2 
 \ . 
 \eeq
Note that if $P = p_1 +p_2$, then in the notation of \eqref{nonnull} we have $\xi = \l_1 + i \l_2$ and $\mu = \l_1 - i \l_2$, where 
$p_i := \l_i \l_i$, $i=1,2$. 

\section{One-loop triangle in $D=3$}
In this section we describe the explicit evaluation of the three-mass one-loop triangle integral
in three dimensions. We will use Feynman parameters and a Mellin Barnes representation to perform the integrals. The result is finite but since the Mellin Barnes is singular we will have to introduce an intermediate regulator to perform the calculation.
We have
\beqa
\cI^{\rm 3m}(K_1, K_2, K_3) &:= &  \int\!\! {d^3 l} \, {1 \over( l^2 + i \varepsilon)  ( (l +K_1)^2 + i \varepsilon) ( (l +K_1 + K_2)^2 + i \varepsilon)  }
 \\
& = & -\frac{i\, \pi^2 }{2} \int_{0}^1\!dx \, \int_{0}^{1-x}\!dy\, 
\left[ x(1-x-y) t_1+y(1-x-y) t_2 + x y t_3\right]^{-3/2} \nonumber
\  ,
\eeqa 
where we defined $t_i := -K_i^2 - i \varepsilon$. To arrive at the second line we have introduced Feynman parameters $x, y$ and performed the loop integration after Wick rotating
$l^0 \to i l^0$. Next, we break up the denominator into three terms using a double Mellin Barnes integral to arrive at
\beqa\label{inter}
&&-\frac{i\, \pi^2 }{2\pi \Gamma(3/2) (2\pi i)^2} \int_{-i \infty}^{+i \infty} dz dw \, \Gamma(-z) \Gamma(-w) \Gamma\Big(\frac{3}{2}+z+w\Big) t_1^{-\frac{3}{2}-z-w} t_2^z \, t_3^w \nonumber\\
& & \times \int_{0}^1\!dx \, \int_{0}^{1-x}\!dy \, \big[x (1-x-y)\big]^{-\frac{3}{2}+\epsilon} \Big(\frac{y}{x}\Big)^w
\Big(\frac{y}{1-x-y}\Big)^z 
\, ,
\eeqa
where in the last line we have introduced an intermediate regulator, $\epsilon$, for the otherwise ill-defined Feynman parameter integral. The last line of \eqref{inter} integrates then to
\be
\frac{\Gamma(-\frac{1}{2}+\epsilon-w) \Gamma(-\frac{1}{2}+\epsilon-z) \Gamma(1+w+z)}{\Gamma(2 \epsilon)} \ .
\ee
The remaining contour integrals can be evaluated directly or using the Mathematica package
{\tt MB.m} \cite{czakon} which in this case reduce to a double residue at $z=w=-1/2$.
The final result including all factors in the limit $\epsilon \to 0$ is
\be
\label{3masstri}
\cI^{\rm 3m}(K_1, K_2, K_3)={- i \, \pi^3    \over \sqrt{ -K_1^2- i \varepsilon}   \sqrt{ -K_2^2 - i \varepsilon}  \sqrt{ -K_3^2 - i \varepsilon} } \ .
\ee
Let us conclude with a short comment on dual conformal symmetry. Obviously $\cI^{\rm 3m}$
is not invariant under dual conformal inversions. To see this we rewrite momenta in terms
of dual momenta as $K_i = x_i - x_{i+1}$ and denote the internal loop momentum by $x_0$. Then
the three-mass triangle can be written as
\be
\int\!\!\frac{d^3 x_0}{x_{10}^2 x_{20}^2 x_{30}^2} \ ,
\ee
with $x_{ij} \equiv x_i-x_j$. This is obviously non-invariant since under an inversion $x^\mu \to x^\m/x^2$ 
it picks up the factor $x_1^2 x_2^2 x_3^2$. This can be compensated by an appropriate prefactor $\sqrt{-x_{12}^2}
\sqrt{-x_{23}^2} \sqrt{-x_{31}^2} = \sqrt{ -K_1^2 - i \varepsilon} \sqrt{ -K_2^2 - i \varepsilon}  \sqrt{ -K_3^2 - i \varepsilon}$. But this implies  
that the properly normalised three-mass triangle in $D=3$ is a constant. This  is consistent with the fact that it is impossible to write a dual conformal cross-ratio with only three momenta.


\section{The tree-level eight-point amplitude}
In this section we present the calculation for tree-level eight-point amplitude by applying BCFW recursion relations explicitly. Let us start with the recursive diagram in Figure \ref{8-pt-rec}(a), $\cR^{(a)}_{12; 4}$, which can be evaluated by using the four- and six-point results, (\ref{fourtree}) and (\ref{sixtree}),  
\beq
\cR^{(a)}_{12; 4} =
{1\over \sqrt{ P^2_{58} P^2_{34}  } }\,  \int\!\!d^3 \eta_{\hat{P}_a} \   {1 \over z_{a_1} - z_{a_1}^{-1}  }  \Big[\cM_R( \bar{3}, 4, - \bar{\hat{P}}_a , \hat{2}) \, \cM_L(\bar{5}, 6 , \overline{7} ,8, \bar{\hat{1}}, \hat{P}_a) \Big]_{z = z_{a_1}} 
- \, (z_{a_1} \rightarrow  z_{a_2})  \, .
 \eeq    
Evaluating this, one obtains 
\beqa
\label{a8}
\cR^{(a)}_{12; 4} &:=&   Y^{(1)}_{12;4} \, + \, Y^{(2)}_{12;4}
\nonumber \\
&=&
i{1 \over \sqrt{ P^2_{58} P^2_{34}  }}\,  {1 \over z_{a_1} - z_{a_1}^{-1}  }  
  {\delta^{(3)}(P) \delta^{(6)}(Q) \delta^{(3)}(\sigma_a) \over P^2_{68} \lan 3 \, 4\ran^4  \lan 4 \, \hat{P}_a \ran} 
  \nonumber \\ \cr
  & & \times  
  \bigg[ { \delta^{(3)}(\gamma_{a_1})  
 \over (\lan 6| P_{78}| \hat{1} \ran + i \lan 7 \, 8\ran \lan \hat{P}_a \, 5\ran)
 (\lan 5| P_{67}| 8 \ran + i \lan 6 \, 7\ran \lan \hat{1} \, \hat{P}_a \ran)  } 
  \nonumber \\ \cr
& & +   
{ \delta^{(3)}(\gamma_{a_2})  
 \over (\lan 6| P_{78}| \hat{1} \ran - i \lan 7 \, 8\ran \lan \hat{P}_a \, 5\ran)
 (\lan 5| P_{67}| 8 \ran - i \lan 6 \, 7\ran \lan \hat{1} \, \hat{P}_a \ran) }  \bigg]_{z = z_{a_1}}   
 \nonumber \\ \cr
& &-
 (z_{a_1} \rightarrow  z_{a_2}) \, ,
\eeqa
where the arguments of the fermionic delta functions are 
\beqa
\sigma_a = \lan 3 \, 4\ran \hat{\eta}_{2} + \lan 4 \, \hat{2} \ran \eta_{3} + \lan \hat{2} \, 3\ran \eta_{4} \, ,
\eeqa
and
\beqa
\gamma_{a_{1,2}} = \lan \hat{2} \, \hat{P}_a \ran \epsilon_{ijk} \lan j \, k \ran \eta_i  \mp 
i  \left( \lan \hat{1}| P_{34}| \hat{2} \ran \eta_5 - \lan \hat{2}| P_{34}| 5 \ran \hat{\eta}_{1} + 
\lan 5 \, \hat{1} \ran (\lan 3 \, \hat{2} \ran \eta_3 +  \lan 4 \, \hat{2} \ran \eta_4 ) \right) \, ,
\eeqa
with $i, j, k = 6,7,8,$ and $\hat{P}_a =\hat {p}_{2} + p_3 + p_4$. Finally the on-shell solutions $z_{a_i}$ can be found from  (\ref{explsol}) or (\ref{explsol2}) with  
\beqa 
K_1= P_{58} \,  , \qquad  K_2 = P_{34} \, ,
\eeqa
and similarly one can determine $z_{b_i}$, the on-shell solution of the contribution from Figure \ref{8-pt-rec}(b), which we will consider in the following. For the contribution of Figure \ref{8-pt-rec}(b), we have
\beqa
\label{b8}
\cR^{(b)}_{12; 4} &=&
-{i \over \sqrt{ P^2_{36} P^2_{78}  }}\,  {1 \over z_{b_1} - z_{b_1}^{-1}  }  
  {\delta^{(3)}(P) \delta^{(6)}(Q) \delta^{(3)}(\sigma_b) \over P^2_{46} \lan 7 \, 8 \ran^4 \lan 8 \, \hat{1} \ran} 
  \nonumber \\ \cr
& &\times 
  \bigg[ { \delta^{(3)}(\gamma_{b_1}) 
 \over (\lan 4| P_{56}| \hat{P}_b \ran +  \lan 5 \, 6\ran \lan \hat{2} \, 3\ran)
 (\lan 3| P_{45}| 6 \ran - \lan 4 \, 5\ran \lan \hat{P}_b \, \hat{2}\ran) } 
  \,  \cr
& & +  
{ \delta^{(3)}(\gamma_{b_2}) 
 \over (\lan 4| P_{56}| \hat{P}_b \ran -  \lan 5 \, 6\ran \lan \hat{2} \, 3\ran)
 (\lan 3| P_{45}| 6 \ran + \lan 4 \, 5\ran \lan \hat{P}_b \, \hat{2}\ran) } 
   \bigg]_{z = z_{b_1}} 
   \nonumber \\ \cr
& & -
 (z_{b_1} \rightarrow  z_{b_2}) \, , 
\eeqa
where 
\beqa
\sigma_b = \lan 7 \, 8\ran \hat{\eta}_{1} + \lan 8 \, \hat{1} \ran \eta_{7} + \lan \hat{1} \, 7\ran \eta_{8} \, ,
\eeqa
and
\beqa
\gamma_{b_{1,2}} = \lan \hat{1} \, \hat{P}_b \ran \epsilon_{ijk} \lan j \, k \ran \eta_i  \mp 
 \, \left( \lan \hat{1}| P_{78}| \hat{2} \ran \eta_{3} - \lan \hat{1}| P_{78}| 3 \ran \hat{\eta}_{2} + 
\lan \hat{2} \, 3 \ran (\lan \hat{1} \, 7 \ran \eta_7 +  \lan \hat{1} \, 8 \ran \eta_8 ) \right) \, , 
\eeqa
with $i, j, k = 4, 5, 6,$ and $\hat{P}_{b} = - (p_7 + p_8 + \hat{p}_{1}).$

In summary, the complete tree-level eight-point amplitude is given by the sum of the two contributions \eqref{a8} and \eqref{b8}, 
\beqa
\cM_{\rm tree}(\bar{1},2, \ldots ,\bar{7},  8)\  = \ \cR^{(a)}_{12; 4} \, + \, \cR^{(b)}_{12; 4} \, .
\eeqa


\end{document}